\documentclass[apj]{emulateapj}

\usepackage{url,amsfonts,epsfig}
\usepackage{graphicx}
\usepackage{amsmath}
\usepackage{amssymb}
\usepackage[section]{placeins}
\usepackage{chngpage}
\usepackage{calc}
\usepackage{subfigure}
\usepackage{mathtools}
\usepackage{appendix}
\usepackage{natbib}
\makeatletter 
\renewcommand\@biblabel[1]{}

\shorttitle{Orbital evolution of mass-transferring eccentric binary systems}
\shortauthors{Dosopoulou and Kalogera}
\begin{document}
\def\gap{\;\rlap{\lower 2.5pt
\hbox{$\sim$}}\raise 1.5pt\hbox{$>$}\;}
\def\lap{\;\rlap{\lower 2.5pt
 \hbox{$\sim$}}\raise 1.5pt\hbox{$<$}\;}

\newcommand\sbh{MBH}
\newcommand\NSC{NSC}
\title{Orbital evolution of mass-transferring eccentric binary systems.\\ I. Phase-dependent evolution}

\author{Fani Dosopoulou and Vicky Kalogera}
\email{FaniDosopoulou2012@u.northwestern.edu}
\affil{Center for Interdisciplinary Exploration and Research in Astrophysics (CIERA) and Department of Physics and Astrophysics,
Northwestern University, Evanston, IL 60208}

\begin{abstract}
Observations reveal that mass-transferring binary systems may have non-zero orbital eccentricities. The  time-evolution of the orbital semi-major axis and eccentricity of mass-transferring eccentric binary systems is an important part of binary evolution theory and has been widely studied. However, various different approaches and assumptions on the subject have made the literature difficult to comprehend and comparisons between different orbital element time-evolution equations not easy to make. Consequently, no self-consistent treatment of this phase has been ever included in binary population synthesis codes. In this paper, we present a general formalism to derive the time-evolution equations of the binary orbital elements, treating mass-loss and mass-transfer as perturbations to the general two-body problem. We present the self-consistent form of the perturbing acceleration and the phase-dependent time-evolution equations for the orbital elements under different mass-loss/transfer processes. First, we study the cases of isotropic and anisotropic wind mass-loss. Then, we proceed with the non-isotropic ejection and accretion in conservative as well as non-conservative manner for both point masses and extended bodies. Comparison of the derived equations with similar work in the literature is made and explanation of the existing discrepancies is provided.
 \end{abstract}
 

\section{Introduction}
In modeling of binary populations, most interacting mass-transferring binary systems are assumed to be circular \citep[e.g.][]{2008ApJS..174..223B}. However, 
observations reveal  that 
 semi-detached 
binary systems might have non-zero orbital eccentricities \citep[e.g.,][]{1999AJ....117..587P,2008A&A...480..797B,2013A&A...559A..54V,2014A&A...564A...1B}. In addition, recent observations indicate that High-Mass X-ray Binaries (HMXB) are eccentric, with relatively high eccentricities in the case of Be/X-ray binaries \citep[e.g.,][]{2005A&AT...24..151R,2015A&ARv..23....2W}. The residual eccentricity that has been observed in mass-transferring binary systems is quite surprising, since tidal dissipation might be expected to circularize the orbit and drive the star towards co-rotation \citep[e.g.,][]{1981A&A....99..126H,1998ApJ...499..853E,2008A&A...480..797B}. This implies that the efficiency of tidal circularization in interacting eccentric binaries is not as high as typically assumed. Previous work has shown that especially for long-period systems and solar-type binaries in open clusters tides are not sufficient to circularize the orbit, leaving the systems with a considerable non-zero eccentricity \citep[e.g.,][]{2013A&A...559A..54V,2012A&A...548A...6V,2015A&A...579A..49V,2005ApJ...620..970M}. Possible mechanisms that have been proposed to help the system retain a finite eccentricity are mass-loss and mass-transfer processes.\par
Mass-loss/transfer can indeed enhance eccentricity \citep[e.g.,][]{2015A&A...579A..49V,2008A&A...480..797B,2000A&A...357..557S} and is a common evolutionary phase in a binary system, being responsible for many observed astrophysical phenomena such as Type Ia supernovae, X-ray binaries, neutron-stars spin up as well as orbital contraction or expansion. Mass can be ejected or accreted in a binary  system in many different and complicated ways and is phase-dependent in an eccentric orbit. Commonly studied cases include isotropic wind mass-loss \citep[e.g.,][]{1963Icar....2..440H}, Roche-Lobe-Overflow (RLOF) \citep[e.g.,][]{2007ApJ...667.1170S,2009ApJ...702.1387S} and Bondi-Hoyle accretion \citep[e.g.,][]{1944MNRAS.104..273B,1988A&A...205..155B,2002MNRAS.329..897H}. All the aforementioned processes change the total mass, energy and orbital angular momentum of the binary, which corresponds to a change in the binary orbital elements. The effect of mass-transfer on the orbital elements of the orbit goes back a while ago \citep[e.g.,][]{1963Icar....2..440H,1956AJ.....61...49H} and continues to recent years with work studying various and different cases of mass-loss and mass-transfer \citep[e.g.,][]{2007ApJ...667.1170S,2009ApJ...702.1387S,2010ApJ...724..546S,2011MNRAS.417.2104V,2013MNRAS.435.2416V}. \par
In the existing literature, to our knowledge, the time-evolution of the orbital elements of a binary undergoing mass-loss and mass-transfer has been derived following two different approaches. One involves calculating time variations of the orbital elements based on changes in the total orbital energy and angular momentum of the system \citep[e.g.,][]{1956AJ.....61...49H,1988A&A...205..155B,2008A&A...480..797B}. The other is treating mass-loss/transfer as a perturbation in the two-body problem and using the latter to calculate the time-evolution of the orbital elements  \citep[e.g.,][]{1963Icar....2..440H,2009ApJ...702.1387S,2011MNRAS.417.2104V,2013MNRAS.435.2416V}. However, consistency between these two different approaches is not always clear and many assumptions that have implicitly been made following either technique are not always clearly stated. Thus, having a general mathematical formalism to describe the orbital evolution of a binary system undergoing mass-transfer of various types is both useful and important in binary evolution theory. In the first part of this paper, we present how to derive the time-evolution equations of the orbital elements based on the Lagrange perturbation formalism (Lagrange $1808a,1808b,1809$) and then apply this to specific cases of mass-loss and mass-transfer in eccentric binaries. \par
A binary system exposed to different kind of perturbations follows an orbit that is continuously evolving in time, i.e., an open orbit. This means that much attention must be paid to the physical interpretation of the orbital elements we assign to the evolving orbit \citep[e.g.,][]{2011MNRAS.417.2104V,2004A&A...415.1187E}. In this work, we refer to the importance of this interpretation which is connected both to the freedom in the choice of orbital elements and the gauge freedom in orbital mechanics.
\par The paper is organized as follows. In Section 2 we present the general formalism we use to describe the perturbed two-body problem. In section 3 we comment on the importance of the choice of reference frame and present the time-evolution equations of the orbital elements in three different reference frames. In sections 4 and 5 we apply this formalism to the cases of isotropic and anisotropic wind mass-loss respectively. In Section 6 we study conservative as well as non-conservative non-isotropic mass ejection and accretion in the system for extended binary components. In Section 7 we present the form the time-evolution equations of the orbital elements reduce to under the point-mass approximation. Throughout all the aforementioned sections, comparison to previous work in the literature that has considered similar cases is made. Finally, an explanation of any existing discrepancies is provided. In Section 8 we summarize our results. \par
The perturbing acceleration is in principle phase-dependent. In this paper, we present the general form of the perturbing acceleration and the orbital element time-evolution equations, keeping this phase-dependence. In a follow-up paper Dosopoulou $\&$ Kalogera 2016b (from now on Paper II), we orbit-average the phase-dependent time-evolution equations of the orbital elements relevant to different types of mass-loss/transfer in the binary under either the assumption of adiabaticity or for delta-function mass-loss/transfer at periastron. 

\section{The perturbed two-body problem}

Two-body systems like binary stars can be exposed to gravitational perturbations from different agents. These include tidal dissipation, relativistic corrections, gravitational-wave radiation, magnetic fields, inertial forces, mass-loss/mass-transfer phenomena and other. All these physical processes act like perturbing forces to the general two-body problem and need a new mathematical way of treatment compared to the unperturbed case. Due to the various perturbations, each body is no longer moving in the actual physical trajectory (conic section) it would if no perturbations existed, but its physical orbit is slowly changing with time. \par
Either called \emph{Variation of constants} in pure mathematical language or \emph{Varying-Conic method} in purely geometric terms, the method that was proposed to study these systems was advanced and completed by Lagrange. This method involves an approximation of the true physical orbit by a family of evolving instantaneous conic sections.\par
In subsection $2.1$, we will prove that the latter representation is not unique, thus one should be very meticulous when applying this mathematical formulation. Specifically, the actual physical orbit of the body is open, the orbital elements derived though this method are not the classic orbital elements of a closed orbit. Rather, they either can have a possible physical interpretation under certain assumptions or they should be treated just as mathematical tools otherwise.We shall discuss this issue in more detail in the next sections.

\subsection{Variation of Constants method}

The general reduced two-body problem is described by the equation
\begin{equation}\label{eq1}
\ddot{\bold{r}}=-\frac{GM}{r^{3}}\bold{r}
 \end{equation}
where $G$ is the gravitational constant, $M$ is the total mass of the system and $r$ is the relative position between the two bodies. The general solution to this problem is a Keplerian conic section defined by six constant orbital elements $\left\lbrace C_{i} \right\rbrace=\left\lbrace C_{1},...,C_{6} \right\rbrace$. \par
In a specified inertial coordinate system, this conic is described by
\begin{equation}\label{eq2}
\bold{r}=\bold{r}(C_{1},...,C_{6},t),\:\:\: \dot{\bold{r}}=(C_{1},...,C_{6},t)
 \end{equation}
where by definition
\begin{equation}\label{eq3}
\bold{v}=\left( \frac{\partial{\bold{r}}}{\partial{t}}\right)_{C_{i}=const.}
 \end{equation}
is the relative velocity between the two bodies in the system.\par
The \emph{perturbed} two-body problem can then be written as
\begin{equation}\label{aa}
\ddot{\bold{r}}=-\frac{GM}{r^{3}}\bold{r}+\bold{F}(\bold{r},\dot{\bold{r}})
 \end{equation}
where the perturbing force $\bold{F}(\bold{r},\dot{\bold{r}})$ depends upon both the relative position and velocity.\par
Equation $(\ref{aa})$ can be solved assuming that at each instant of time, the true orbit can be approximated by an instantaneous Keplerian Conic section which is changing over time through the now \emph{time-dependent} orbital elements $C_{i}$, i.e.,
\begin{equation}\label{eq5}
\bold{r}=\bold{r}(C_{1}(t),...,C_{6}(t),t)
 \end{equation}
and the velocity of the body will now be given by 
\begin{equation}\label{eq6}
\dot{\bold{r}}=\frac{\partial{\bold{r}}}{\partial{t}}+\sum_{i=1}^{6}\frac{\partial{\bold{r}}}{\partial{C_{i}}}\frac{dC_{i}}{dt}.
 \end{equation}
Equations $(\ref{aa})$ and $ (\ref{eq6})$ constitute a system of three independent equations, one independent variable (time $t$) and six dependent variables (orbital elements $C_{i}(t)$). This means that  the parametrization in terms of the orbital elements is not unique and additional constraints need to be applied.

\subsubsection{Lagrange constraint - Osculating elements}

Imposing the \emph{Lagrange constraint}
\begin{equation}\label{eq7}
\sum_{i=1}^{6}\frac{\partial{\bold{r}}}{\partial{C_{i}}}\frac{dC_{i}}{dt}=0
 \end{equation}
makes the orbital elements $C_{i}(t)$ \emph{osculating}. This means that the instantaneous conics are tangential to the perturbed actual physical orbit, as it can be seen by equation $(\ref{eq7})$ and is also depicted in Figure $1$.\par
Another way to think about osculating orbits is that these are the orbits that the body would follow if perturbations were to cease instantaneously. \par
Equations $(\ref{aa})$,$(\ref{eq6})$  and $(\ref{eq7})$ constitute now a well-defined system of six equations and six variables.\par
Following the Lagrange constraint, the perturbed equation of motion $(\ref{aa})$ can be written as
\begin{equation}\label{eq8}
\frac{\partial^{2}\bold{r}}{\partial{t^{2}}}+\frac{GM}{r^{3}}\bold{r}
+\sum_{i=1}^{6}\frac{\partial{\bold{v}}}{\partial{C_{i}}}\frac{dC_{i}}{dt}=\bold{F}
 \end{equation}
where $\bold{r}$ refers to the solution of the unperturbed two-body problem, i.e., satisfies equation $(\ref{eq1})$.
\par This simplifies equation $(\ref{eq8})$ to 
\begin{equation}\label{eq9}
\sum_{i=1}^{6}\frac{\partial{\bold{v}}}{\partial{C_{i}}}\frac{dC_{i}}{dt}=\bold{F}
 \end{equation}
which using the definition of Lagrange and Poisson brackets can be decoupled to read 
\begin{equation}\label{eq10}
\frac{dC_{i}}{dt}=-\sum_{j=1}^{6}\left\lbrace C_{i}C_{j}\right\rbrace\frac{\partial{\bold{r}}}{\partial{C_{j}}}\bold{F}.
 \end{equation}
\par Poisson brackets are defined by 
\begin{equation}\label{eq11}
\left\lbrace C_{k}C_{i}\right\rbrace=
\frac{\partial{C_{k}}}{\partial{\bold{r}}}\frac{\partial{C_{i}}}{\partial{\bold{v}}}-\frac{\partial{C_{k}}}{\partial{\bold{v}}}\frac{\partial{C_{i}}}{\partial{\bold{r}}},
 \end{equation}
they are the negative inverse to Lagrange brackets
\begin{equation}\label{eq12}
\left[ C_{k}C_{i}\right ]=
\frac{\partial{\bold{r}}}{\partial{C_{k}}}\frac{\partial{\bold{v}}}{\partial{C_{i}}}-\frac{\partial{\bold{r}}}{\partial{C_{i}}}\frac{\partial{\bold{v}}}{\partial{C_{k}}},
 \end{equation}
and do not depend on time explicitly.
\par Equations $(\ref{eq10})$ give the time-evolution equations of the six osculating orbital elements, which are invariant with respect to the change of coordinates and/or orbital parametrization.
\par We choose the six Keplerian elements $(a,e,i,\Omega,\omega,\nu)$, namely semi-major axis, eccentricity, inclination, argument of periapsis, longitude of the ascending node, and true anomaly $\nu$. Also, we define $n=(GM/a^{3})^{1/2}$ as the mean motion.\par
Making use of the following Poisson brackets
\begin{align}
\left\lbrace\sigma a\right\rbrace &=-\left\lbrace a \sigma \right\rbrace=\frac{2}{na}\\
\left\lbrace\sigma e\right\rbrace &=\frac{1-e^{2}}{na^{2}e},\left\lbrace\omega e\right\rbrace =-\frac{\sqrt{1-e^{2}}}{na^{2}e}\\
\left\lbrace\sigma a\right\rbrace &=-\frac{\sqrt{1-e^{2}}}{na^{2}e}\\
\left\lbrace\Omega i\right\rbrace &=-\frac{1}{na^{2}\sqrt{1-e^{2}}\sin i}\\
\left\lbrace\omega i\right\rbrace &=-\frac{\cos i}{na^{2}\sqrt{1-e^{2}}\sin i}
\end{align}
yields the following Lagrange time-evolution equations for the aforementioned six orbital elements 
\begin{align}
\frac{da}{dt}&=\frac{2}{na}\bold{F}\frac{\partial{\bold{r}}}{\partial{\sigma}}\label{generala}\\
\frac{de}{dt}&=\frac{1-e^{2}}{na^{2}e}\bold{F}\frac{\partial{\bold{r}}}{\partial{\sigma}}-\frac{\sqrt{1-e^{2}}}{na^{2}e}\bold{F}\frac{\partial{\bold{r}}}{\partial{\omega}}\label{generale}\\
\frac{di}{dt}&=\frac{1}{na^{2}\sqrt{1-e^{2}}\sin i}\left[\cos i\:\:\bold{F}\frac{\partial{\bold{r}}}{\partial{\omega}}
-\bold{F}\frac{\partial{\bold{r}}}{\partial{\Omega}}\right]\label{generali}\\
\frac{d\omega}{dt}&=\frac{\sqrt{1-e^{2}}}{na^{2}e}\bold{F}\frac{\partial{\bold{r}}}{\partial{e}}-\frac{\cos i}{na^{2}\sqrt{1-e^{2}}\sin i}\bold{F}\frac{\partial{\bold{r}}}{\partial{i}}.\label{generalomega}\\
\frac{d\Omega}{dt}&=\frac{1}{na^{2}\sqrt{1-e^{2}}\sin i}\bold{F}\frac{\partial{\bold{r}}}{\partial{i}}\label{generalOmega}\\
\frac{d\nu}{dt}&=\frac{n (1+e\cos \nu)^{2}}{(1-e^{2})^{3/2}}-\frac{d\omega}{dt}-\cos i \frac{d\Omega}{dt}.\label{generalsigma}
\end{align}
\par We note here that the orbital elements appearing in equations $(\ref{generala})$-$(\ref{generalsigma})$ are, as we will explain in detail in the next sections, the so-called \emph{osculating elements}. The physical meaning of these elements as well as their uniqueness with respect to any other choice of orbital elements will be explained in the next sections as well.
 \par Equations $(\ref{generala})$-$(\ref{generalsigma})$ describe the time-evolution of all the six orbital elements. Because in this paper we are interested only in perturbations induced by mass-loss/transfer processes and for reasons we describe in detail in the last paragraph of the following section, in this paper we present only the time-evolution of the semi-major axis $a$ and the eccentricity $e$ [see equations $(\ref{generala})$ and $(\ref{generale})$]. However, as we can see from equations $(\ref{generalomega})$-$(\ref{generalsigma})$ for a general perturbation the orbit undergoes various precessions. At the end of the following section we discuss briefly these precessions as well. In Paper II we discuss in more detail the importance of these precessions within the context of secular evolution.
\subsubsection{Introducing Gauge freedom in orbital mechanics - Non-osculating elements}
\par According to \citet{2004A&A...415.1187E}, imposing the Lagrange constraint is not necessary and one is free to set 
\begin{equation}\label{eq18}
\sum_{i=1}^{6}\frac{\partial{\bold{r}}}{\partial{C_{i}}}\frac{dC_{i}}{dt}=\bold{\Phi}
 \end{equation}
where $\bold{\Phi}$ is called \emph{the gauge} and represents the gauge freedom in orbital mechanics.
\par Generalizing more the formulation by writing the perturbing force 
$\bold{F}$ in terms of a perturbing Lagrangian 
\begin{equation}
\bold{F}=\frac{\partial{\Delta L}}{\partial{\bold{r}}}
-\frac{d}{dt}\left(\frac{\partial{\Delta L}}{\partial{\dot{\bold{r}}}}\right)\label{forcelag}
\end{equation}
one can re-write equations $(\ref{generala})$ and $(\ref{generale})$ as follows \citep[e.g.,][]{2004A&A...415.1187E}
\begin{align}
\begin{split}
\frac{da}{dt}&=\frac{2}{na}\left[
\frac{\partial{(-\Delta H)}}{\partial{\sigma}}
-\frac{\partial{\Delta L}}{{\partial{\dot{\bold{r}}}}}\frac{\partial}{\partial{\sigma}}\left( \bold{\Phi} +\frac{\partial{\Delta L}}{{\partial{\dot{\bold{r}}}}}\right)\right.\\
&\left. -\left( \bold{\Phi} +\frac{\partial{\Delta L}}{{\partial{\dot{\bold{r}}}}}\right)\frac{\partial{\bold{g}}}{\partial{\sigma}}
-\frac{\partial{\bold{f}}}{\partial{\sigma}}\frac{d}{dt}\left( \bold{\Phi} +\frac{\partial{\Delta L}}{{\partial{\dot{\bold{r}}}}}\right)\right]
\end{split}\label{hugea}\\
\begin{split}
\frac{de}{dt}&=\frac{1-e^{2}}{na^{2}e}\left[
\frac{\partial{(-\Delta H)}}{\partial{\sigma}}
-\frac{\partial{\Delta L}}{{\partial{\dot{\bold{r}}}}}\frac{\partial}{\partial{a}}\left( \bold{\Phi} +\frac{\partial{\Delta L}}{{\partial{\dot{\bold{r}}}}}\right)\right.\\
&\left. -\left( \bold{\Phi} +\frac{\partial{\Delta L}}{{\partial{\dot{\bold{r}}}}}\right)\frac{\partial{\bold{g}}}{\partial{\sigma}}
-\frac{\partial{\bold{f}}}{\partial{\sigma}}\frac{d}{dt}\left( \bold{\Phi} +\frac{\partial{\Delta L}}{{\partial{\dot{\bold{r}}}}}\right)\right]
\end{split}\label{hugee}\\
\begin{split}
&\nonumber -\frac{\sqrt{1-e^{2}}}{na^{2}e}\left[
\frac{\partial{(-\Delta H)}}{\partial{\omega}}
-\frac{\partial{\Delta L}}{{\partial{\dot{\bold{r}}}}}\frac{\partial}{\partial{\omega}}\left( \bold{\Phi} +\frac{\partial{\Delta L}}{{\partial{\dot{\bold{r}}}}}\right)\right. \\
&\left.-\left( \bold{\Phi} +\frac{\partial{\Delta L}}{{\partial{\dot{\bold{r}}}}}\right)\frac{\partial{\bold{g}}}{\partial{\omega}}
-\frac{\partial{\bold{f}}}{\partial{\omega}}\frac{d}{dt}\left( \bold{\Phi} +\frac{\partial{\Delta L}}{{\partial{\dot{\bold{r}}}}}\right)\right]
\end{split}
\end{align}
where $\Delta H$ is the perturbing Hamiltonian defined by 
\begin{equation}
\Delta H= -\left[ \Delta L+\frac{1}{2}\left(\frac{\partial \Delta L}{\partial \bold{\dot{r}}}\right)^{2}\right]
\end{equation}
and functions $\bold{f}$ and $\bold{g}$ have the same functional form as the unperturbed-case position and velocity vector but are now time dependent through the time-dependent orbital elements $C_{i}$. We note here that equation $(\ref{hugee})$ seems to have a singularity in the case of circular orbits ($e=0$). This singularity is not real since to take the limit of equation $(\ref{hugee})$ for $e \rightarrow 0$ one needs first to calculate explicitly the form of the partial derivatives included in this equation as functions of the orbital elements.\par
For a general gauge $\bold{\Phi}$ they are connected to the perturbed velocity and position by the relations
\begin{align}
\bold{r}&=\bold{f}(C_{1},...,C_{6},t)\\
\dot{\bold{r}}&=\frac{\partial{\bold{f}}}{\partial{t}}+\bold{\Phi}\label{gauge1}\\
&=\bold{g}+\bold{\Phi}.\label{gauge2}
\end{align}
\par Equations $(\ref{gauge1})$ and $(\ref{gauge2})$ constitute a  physical interpretation of the gauge freedom.\par
Choosing the Lagrange constraint $\bold{\Phi}=0$, one enforces the family of the approximating conics to be always tangent to the physical orbit of the body. In other words, this means that the orbital elements under this constraint are osculating elements and describe the conic the body would follow if the perturbation ceased instantaneously.\par
Any other non-zero constraint $\bold{\Phi}$ would lead to orbital elements that do not describe conics that are tangential to the real orbit, but differ from the tangent velocity vector by $\bold{\Phi}$ [see Figure 2]. These elements are the so-called \emph{non-osculating} elements 
and can be used in some cases as useful mathematical tools to describe the problem \citep[e.g.,][]{2002ima..rept....1E}.\par
It is obvious that a real physical interpretation can be assigned only to the osculating elements (Lagrange constraint), since only those describe conics that are always and at any time tangent to the actual path of the body. \par
We should point out here that even the osculating orbital elements derived by this formulation are mathematical tools that can be used to approximately describe the real trajectory of the body. This means that one should be very meticulous when interpreting them. The physical time-evolution of the system is precisely expressed with the time-derivative of the position vector of the body $\bold{\dot{r}}$.\par
The time-evolution equations for the osculating orbital elements include periodic terms that describe the natural oscillations these elements undergo due to the periodicity of the orbit. When the change in the orbital elements over one orbit  is small, then an orbit-averaged method can be adopted to smooth out these oscillations leading to the secular (orbit-averaged) osculating orbital elements. In addition, if the perturbation timescale is much longer than the orbital timescale then these secular time-evolution equations can be resolved into a simpler form. The regime where a perturbation is changing slowly throughout the orbit so that it can be considered constant over an orbit (i.e., an ``adiabatic invariant") is called the adiabatic regime. In this regime the ``adiabatic" secular time-evolution equations for the osculating orbital elements are valid. When the orbital period becomes gradually comparable to the perturbation timescale, then adiabaticity is broken and the ``adiabatic" secular time-evolution equations no longer apply. \citet{2011MNRAS.417.2104V} studied the oscillations of the osculating orbital elements in the case of mass-loss in the adiabatic regime as well as in the regime following the breaking of adiabaticity.\par
When applying equations $(\ref{generala})$ and $(\ref{generale})$ or alternatively $(\ref{hugea})$ and $(\ref{hugee})$ to any kind of problem, one needs to know the various partial derivatives of  $\bold{f}$ and $\bold{g}$ with respect to the orbital elements. For this reason, we calculated and we present in Appendix A all the needed mathematical quantities as functions of the true anomaly $\nu$.
\begin{figure}
  \centering
    \includegraphics[width=0.5\textwidth]{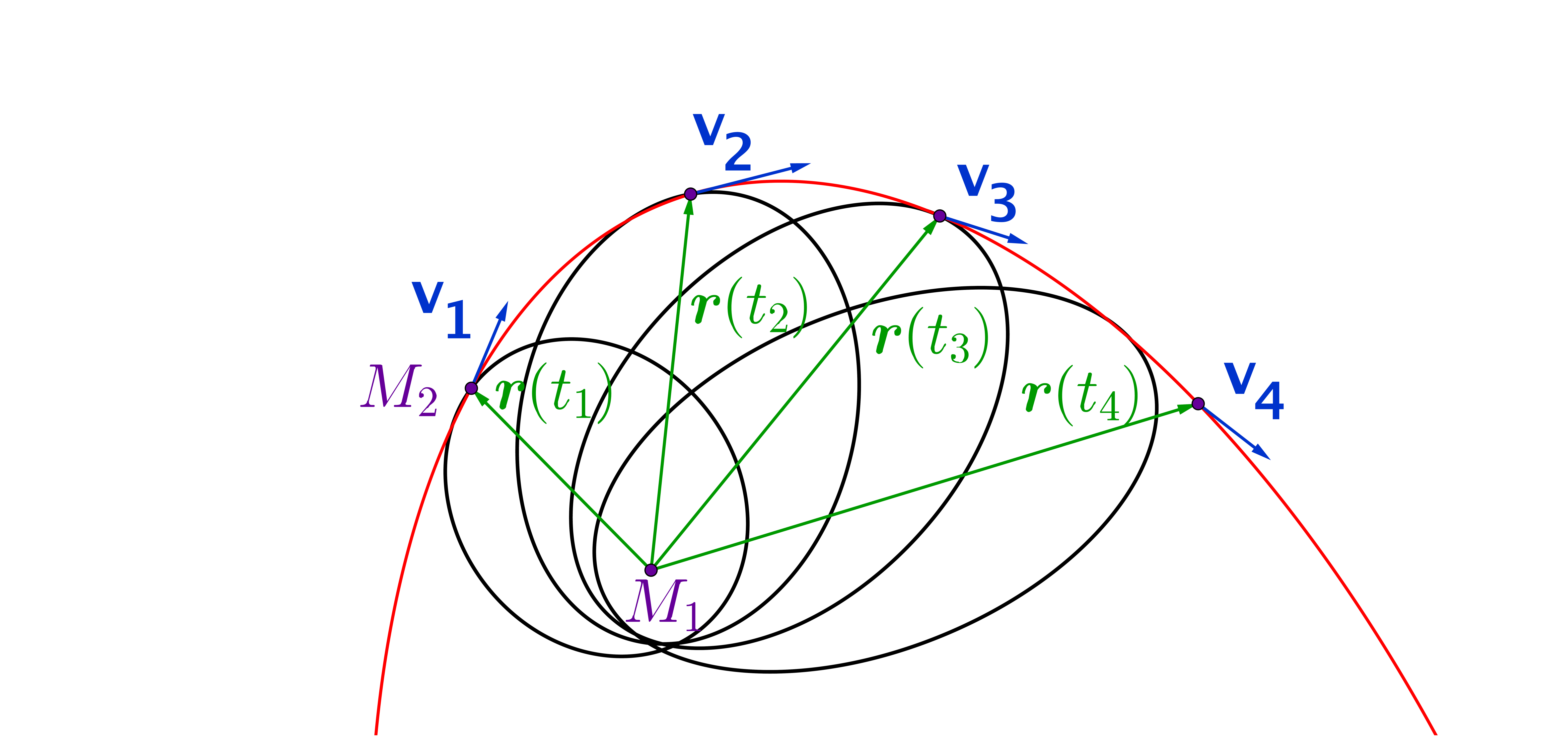}
     \caption{Definition of osculating elements. The instantaneous conics ale always tangential to the perturbed actual physical orbit of the body.}
\end{figure}
\begin{figure}
  \centering
    \includegraphics[width=0.5\textwidth]{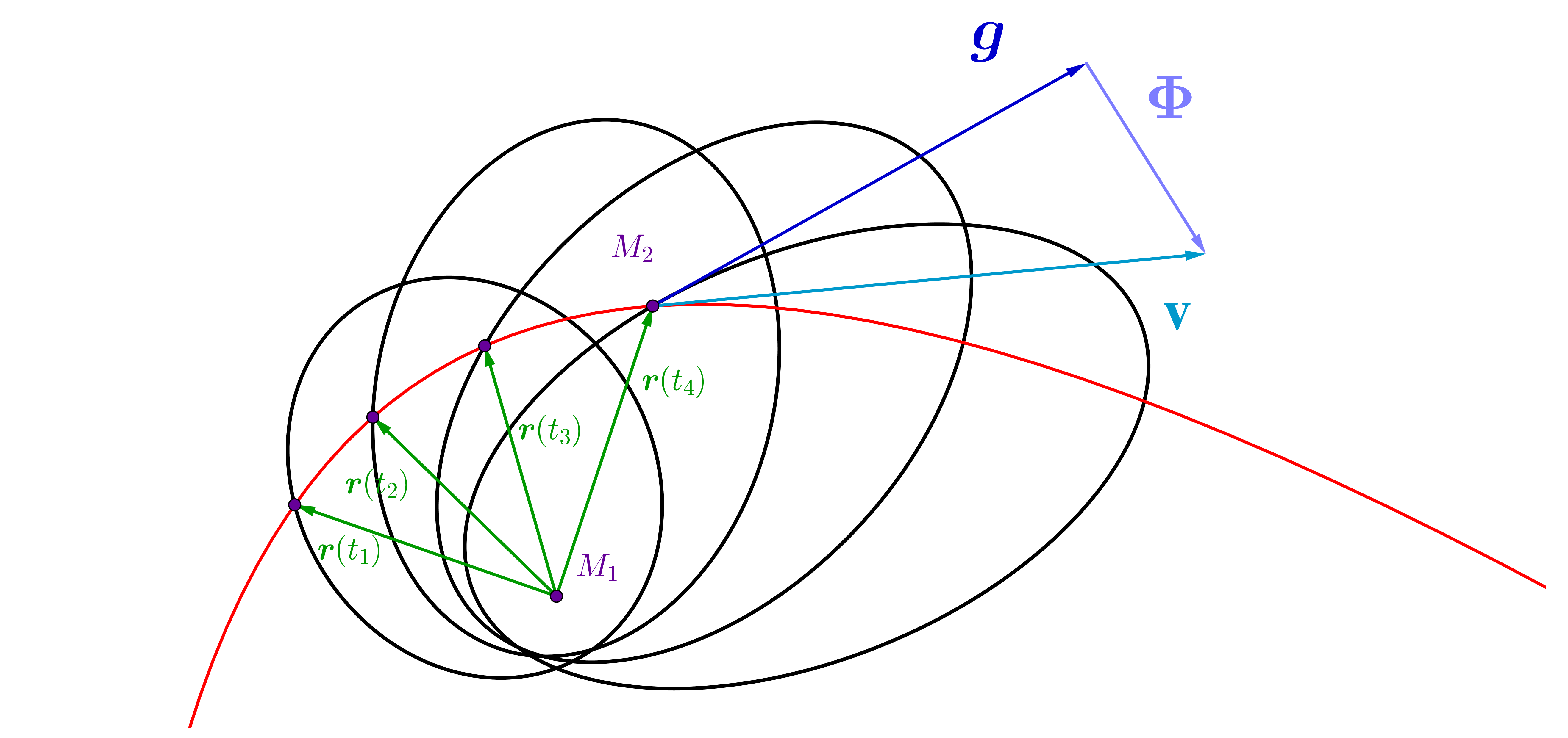}
  \caption{Definition of non-osculating elements. The tangent velocity $\bold{g}$ to the instantaneous conics differs from the tangent velocity vector to the actual physical orbit of the body $\bold{v}=\bold{\dot{r}}$ by the gauge $\bold{\Phi}$. The latter introduces the gauge freedom in orbital mechanics.}
\end{figure}
\par Before proceeding to the application of the zero-gauge and gauge-free aforementioned formulations to specific physical problems, we would like to make a comment on the usefulness and advantages of each of them.\par
Based on what we presented here there are two ways one can derive the time-evolution equations of the orbital elements. One is using the perturbing force $\bold{F}$ directly applying the osculating equations $(\ref{generala})$ and $(\ref{generale})$ and the other is using the equivalent Lagrangian $\Delta L$ defined by equation $(\ref{forcelag})$ and applying equations $(\ref{hugea})$ and $(\ref{hugee})$. Although these two different approaches are completely equivalent eventually leading to the same equations for the osculating orbital elements, the second method which makes use of the Lagrangian is more general because it introduces the gauge freedom. This means that the latter is the only method one could use to choose a gauge different than zero. There are cases where the use of a gauge other than zero is needed, e.g., when the Delaunay variables need to remain canonical for a general perturbation depending both on positions and velocities of the bodies \citep[e.g.,][]{2003JMP....44.5958E,2004A&A...415.1187E}. \par
In the next sections, we will apply the gauge-free method to derive and compare the orbital element time-evolution equations between a zero-gauge and a different gauge $\bold{\Phi}$, called the \emph{generalized Lagrange constraint}, which is commonly used in the literature \citep[e.g.,][]{2002ima..rept....1E,2008AdSpR..42.1313G} and is defined by
\begin{equation}
\bold{\Phi}=-\frac{\partial \Delta L}{\partial \bold{\dot{r}}}.\label{nonoscgauge}
\end{equation}
\par Using this gauge in equations $(\ref{hugea})$ and $(\ref{hugee})$ gives the time-evolution equations of the so-called \emph{contact} orbital elements. As mentioned above, although equations $(\ref{hugea})$ and $(\ref{hugee})$ are simplified by using this gauge, the derived contact orbital elements have no physical meaning compared to the zero-gauge and the relevant osculating orbital elements.
\section{The importance of reference frame choice}
\par It is important to note that equations $(\ref{generala})$ and $(\ref{generale})$ are in vector form and thus independent of the reference frame used in
each specific problem. After decomposing the perturbing force $\bold{F}$ in these equations into its components in a chosen reference frame, one should be careful in interpreting the
resulting equations for the semi-major axis and the eccentricity. In this section we discuss the importance of the reference frame choice pointing out that a vanishing component of the perturbing force in one specific reference frame does not guarantee any other vanishing component of the perturbing force in a different reference frame [see the rotation matrices $(\ref{trans1})$-$(\ref{trans3})$].\par
The most commonly used reference frames $K_{I},K_{J},K_{R}$ to describe a two-body problem are depicted in Figure $3$. Reference system $K_{I}({\bold{\hat{x}},\bold{\hat{y}},\bold{\hat{z}}})$ is the inertial reference system. In the reference system $K_{J}({\bold{\hat{p}},\bold{\hat{q}},\bold{\hat{l}}})$, the unit vector $\bold{\hat{p}}$ is along the eccentricity vector and points always into the direction of the periastron (rotated by the two angles $i$ and $\omega$ relative to $K_{I}$). Finally, in the reference system $K_{R}({\bold{\hat{r}},\bold{\hat{\tau}},\bold{\hat{n}}})$, the unit vector $\bold{\hat{r}}$ is along the relative position
vector between the two bodies in the binary (rotated relative to $K_{J}$ by the angle of true anomaly $\nu$).
\par Acceleration $\bold{A}$ can be written in these reference frames as follows 
\begin{align}
\bold{A}&=A_{x}\bold{\hat{x}}+A_{y}\bold{\hat{y}}+A_{z}\bold{\hat{z}}\label{in}\\
\bold{A}&=A_{p}\bold{\hat{p}}+A_{q}\bold{\hat{q}}+A_{l}\bold{\hat{l}}
\label{ref1}\\
\bold{A}&=A_{r}\bold{\hat{r}}+A_{\tau}\bold{\hat{\tau}}+A_{n}\bold{\hat{z}}.\label{ref2}
\end{align}
The form of equations $(\ref{generala})$ and $(\ref{generale})$ will be different based on the reference frame chosen each time in the problem. The use of the correct set of equations for the chosen reference frame is important since using the incorrect set of equations can lead to misinterpretation of the system's orbital evolution.Here we present this set of equations for three different refernce frames.
\par For decomposition $(\ref{in})$ in the reference system $K_{I}({\bold{\hat{x}},\bold{\hat{y}},\bold{\hat{z}}})$ equations $(\ref{generala})$ and $(\ref{generale})$ take the form
\begin{align}
\begin{split}
\label{in1}\left(\frac{da}{dt}\right)_{\bold{\hat{x}},\bold{\hat{y}},\bold{\hat{z}}}&=\frac{2}{n\sqrt{1-e^{2}}}\left[\left\lbrace
C_{1}\cos{i}\sin{\Omega}+C_{2}\cos{\Omega}\right\rbrace A_{x}
\right.\\
&\left. -\left\lbrace
C_{1}\cos{i}\cos{\Omega}-C_{2}\sin{\Omega}\right\rbrace A_{y}\right.\\
&\left. -\left\lbrace C_{1}\sin{i}\right\rbrace A_{z}\right]
\end{split}
\\
\begin{split}
\label{in2}\left(\frac{de}{dt}\right)_{\bold{\hat{x}},\bold{\hat{y}},\bold{\hat{z}}}&=\frac{\sqrt{1-e^{2}}}{2an(1+e\cos{\nu})}
\left[ 
\left\lbrace
C_{6}\cos{i}\sin{\Omega}+
C_{5}\cos{\Omega}\right\rbrace A_{x}
\right. \\
&\left. -\left\lbrace
C_{6}\cos{i}\cos{\Omega}-C_{5}\sin{\Omega}\right\rbrace A_{y}
\right. \\
&\left.-\left\lbrace
C_{6}\sin{i}\right\rbrace A_{z}\right]
\end{split}
\end{align}
where the different $C_{i}$ are defined by
\begin{align}
C_{1}&=e\cos{\omega}+\cos (\nu+\omega)\\
C_{2}&=e\sin{\omega}+\sin (\nu+\omega)\\
C_{5}&=(2e\cos{\nu}+1+\cos^{2}\nu)\sin{\omega}\\
&+2(e+\cos{\nu})\cos{\omega}\sin{\nu}\\
C_{6}&=(2e\cos{\nu}+1+\cos^{2}\nu)\cos{\omega}\\
&-2(e+\cos{\nu})\sin{\omega}\sin{\nu}.
\end{align}
\par For decomposition $(\ref{ref1})$ in the reference system $K_{J}({\bold{\hat{p}},\bold{\hat{q}},\bold{\hat{l}}})$ equations $(\ref{generala})$ and $(\ref{generale})$ become
\begin{align}
\begin{split}
\label{ref11}\left(\frac{da}{dt}\right)_{\bold{\hat{p}},\bold{\hat{q}},\bold{\hat{l}}}&=\frac{2}{n\sqrt{1-e^{2}}}\left[-A_{p}\sin{\nu}
\right. \\&\left. +A_{q}(e+\cos{\nu})\right]
\end{split}\\
\begin{split}
\label{ref12}\left(\frac{de}{dt}\right)_{\bold{\hat{p}},\bold{\hat{q}},\bold{\hat{l}}}&=\frac{\sqrt{1-e^{2}}}{na(1+e\cos{\nu})}
\left[-A_{p}\sin{\nu}(e+\cos{\nu})
\right. \\
&\left. +A_{q}(2e\cos{\nu}+1+\cos^{2}\nu)\right]
\end{split}
\end{align}
\par For decomposition $(\ref{ref2})$ in the reference system $K_{R}({\bold{\hat{r}},\bold{\hat{\tau}},\bold{\hat{n}}})$ equations $(\ref{generala})$ and $(\ref{generale})$ yield
\begin{align}
\begin{split}
\label{ref21}\left(\frac{da}{dt}\right)_{\bold{\hat{r}},\bold{\hat{\tau}},\bold{\hat{n}}}&=\frac{2}{n\sqrt{1-e^{2}}}\left[A_{r}e\sin{\nu}
\right. \\ & \left. +A_{\tau}(1+e\cos{\nu})\right]
\end{split}\\
\begin{split}
\label{ref22}\left(\frac{de}{dt}\right)_{\bold{\hat{r}},\bold{\hat{\tau}},\bold{\hat{n}}}&=\frac{(1-e^{2})^{1/2}}{na}\left[A_{r}\sin{\nu}
\right. \\
&\left. +A_{\tau}\frac{2\cos{\nu}+e(1+\cos^{2}\nu)}{1+e\cos{\nu}}\right]
\end{split}
\end{align}
\par Transformation between the different reference systems is done using the appropriate rotation matrices. Specifically, the cyclic transformation $K_{I}\rightarrow K_{J}\rightarrow K_{R}$ is done using the relations
\begin{align}
&Q^T{}K_{I}\rightarrow K_{J}\label{trans1} \\
&R K_{J}\rightarrow K_{R}\label{trans2}\\
&Q^{T}K_{I}\rightarrow K_{R},\omega\rightarrow \omega+\nu\label{trans3}
\end{align} 
where the right arrow in equation $(\ref{trans3})$ means substituting $\omega$ with $\omega + \nu$ in the rotation matrix $Q^{T}$. The rotation matrices $\bold{Q}$ and $\bold{R}$ are defined by
 \begin{equation}
 \bold{Q} = 
 \begin{pmatrix}
  Q_{11} & Q_{12}  & Q_{13} \\
  Q_{21} & Q_{22} &  Q_{23} \\
  Q_{31} & Q_{32}  & Q_{33} 
 \end{pmatrix}\label{Q}
\end{equation}
and 
\begin{align}
Q_{11}&=\cos{\Omega}\cos{\omega}-\sin{\Omega}\sin{\omega}\cos{i}\\
Q_{12}&=-\cos{\Omega}\sin{\omega}-\sin{\Omega}\cos{\omega}\cos{i}\\
Q_{13}&=\sin{\Omega}\sin{i}\\
Q_{21}&=\sin{\Omega}\cos{\omega}-\cos{\Omega}\sin{\omega}\cos{i}\\
Q_{22}&=-\sin{\Omega}\sin{\omega}-\cos{\Omega}\cos{\omega}\cos{i}\\
Q_{23}&=-\cos{\Omega}\sin{i}\\
Q_{31}&=\sin{\omega}\sin{i}\\
Q_{32}&=\cos{\omega}\sin{i}\\
Q_{33}&=\cos{i}
\end{align}
and 
 \begin{equation}
 \bold{R} = 
 \begin{pmatrix}
  \cos{\nu} & \sin{\nu}  & 0\\
  -\sin{\nu}  & \cos{\nu}  &  0 \\
  0& 0  & 1
 \end{pmatrix}.\label{R}
\end{equation}
\par In an attempt to clarify the meaning of equations $(\ref{ref1})$  and  $(\ref{ref2})$ we mention here that the latter do not refer to inertial reference frames. The reference frames $K_J$ and $K_R$ are \emph{not inertial} since they are rotating. Equations $(\ref{ref1})$  and  $(\ref{ref2})$ constitute a decomposition of the perturbing force into its components in the relevant reference frames and are true at any instant of time. The matrices $(\ref{Q})$  and  $(\ref{R})$ are time-dependent since the orbital elements are time-dependent. Thus, the transformations $(\ref{trans1})-(\ref{trans3})$ between the different reference frame components of the force are true at any instant of time. These transformations are constructed in such a way that they include the time-derivative of the base unit vectors in $K_J$ and $K_R$ due to the various precessions ($\dot{\omega},\:\:\dot{\Omega},\:\:\dot{\nu}$). Consequently, the various inertial forces that appear when we transform into a rotating reference frame, namely $K_J$ and $K_R$, are already embedded in the form of the \emph{time-dependent rotation matrices} $(\ref{Q})$ and $(\ref{R})$. Inertial forces like Coriolis acceleration, centripetal acceleration and the Euler acceleration will appear applying the relevant time-dependent rotation matrix to the position and force vectors and substitute the transformed position and force vectors into the form of the equation of motion in the inertial system. It is after this substitution that the inertial forces appear in their known form.
\par We note here that \citet{2013MNRAS.435.2416V} presented equations $(\ref{in1})$ and $(\ref{in2})$ as well as $(\ref{ref11})$ and $(\ref{ref12})$ and were the first to apply them to study the case of anisotropic wind mass-loss [see Section 5 here and Section 5 in Paper II]. However, choosing the reference system $K_{J}$ they incorrectly used in their equations $(35)-(39)$ the components of the perturbation along the inertial Cartesian system $K_{I}$ instead of the components of the perturbation in the reference system $K_{J}$ [compare to our equations $(\ref{ref11})$ and $(\ref{ref12})$]. As previously mentioned, this is important since the incorrect use of the right set of time-evolution equations of the orbital elements for the chosen reference system can lead to a misinterpretation of the evolution changes in the semi-major axis $a$ and eccentricity $e$.
\par As we mentioned before in this paper we are only interested in the time-evolution of the semi-major axis and the eccentricity. The basic reason for this is the type of perturbations induced by mass-loss/transfer processes. To explain this in more detail we choose to decompose equations $(\ref{generala})-(\ref{generalsigma})$ in the reference frame $K_{R}$
\begin{align}
\frac{da}{dt}&=\frac{2}{n\sqrt{1-e^{2}}}\left[F_{r}e\sin \nu + F_{\tau} (1+e\cos \nu)\right]\label{ar}\\
\frac{de}{dt}&=\frac{\sqrt{1-e^{2}}}{na}\left[F_{r}\sin \nu + F_{\tau}\left(\cos \nu + \frac{e+\cos \nu}{1+ e\cos \nu}\right)\right]\label{er}\\
\frac{di}{dt}&=F_{n}\frac{\cos(f+\omega)\sqrt{1-e^{2}}}{na(1+e\cos f)}\label{ir}\\
\frac{d\Omega}{dt}&=\frac{1}{na^{2}\sqrt{1-e^{2}}\sin i}F_{n}\frac{a(1-e^{2})\sin (\nu+\omega)}{1+e\cos \nu}\label{Omegar}\\
\nonumber \frac{d\omega}{dt}&=\frac{\sqrt{1-e^{2}}}{nae}\left(-F_{r}\cos \nu + F_{\tau} \sin \nu \frac{2+e\cos \nu}{1+ e \cos \nu}\right) \\
&-\frac{\cos i}{na\sin i}F_{n}\frac{\sqrt{1-e^{2}}\sin (\nu+\omega)}{1+e\cos \nu}\label{omegar}\\
\frac{d\nu}{dt}&=\frac{n (1+e\cos f)^{2}}{(1-e^{2})^{3/2}}-\frac{d\omega}{dt}-\cos{i}\frac{d\Omega}{dt}\label{sigmar}
\end{align}
As we will see in the following sections the perturbations induced by the mass-loss/transfer processes studied in this paper have the general form $\bold{F} \propto \alpha(\nu)\bold{r}+\beta(\nu)\bold{\dot{r}}$. From equations  $(\ref{ir})$ and $(\ref{Omegar})$ we can see that the inclination $i$ and the longitude of the ascending node $\Omega$ evolve only for a non-zero vertical to the orbital plane component of the perturbing force, i.e., $F_{n} \neq 0$. However, perturbations of the form $\bold{F} \propto \alpha(\nu)\bold{r}+\beta(\nu)\bold{\dot{r}}$ do not contain such a vertical perturbing force component (i.e., $F_{n} =0$) and thus $i$ and $\Omega$ do not evolve in the cases we study here. On the contrary, from equation  $(\ref{omegar})$ we see that the argument of periapsis $\omega$ is indeed precessing due to mass-loss/transfer but we assume this precession is at first order approximation small (i.e., $\dot{\omega} << 1$). Under these considerations, we choose to focus and present only the time-evolution of the semi-major axis $a$ and the eccentricity $e$.
\begin{figure}
  \centering
    \includegraphics[width=0.5\textwidth]{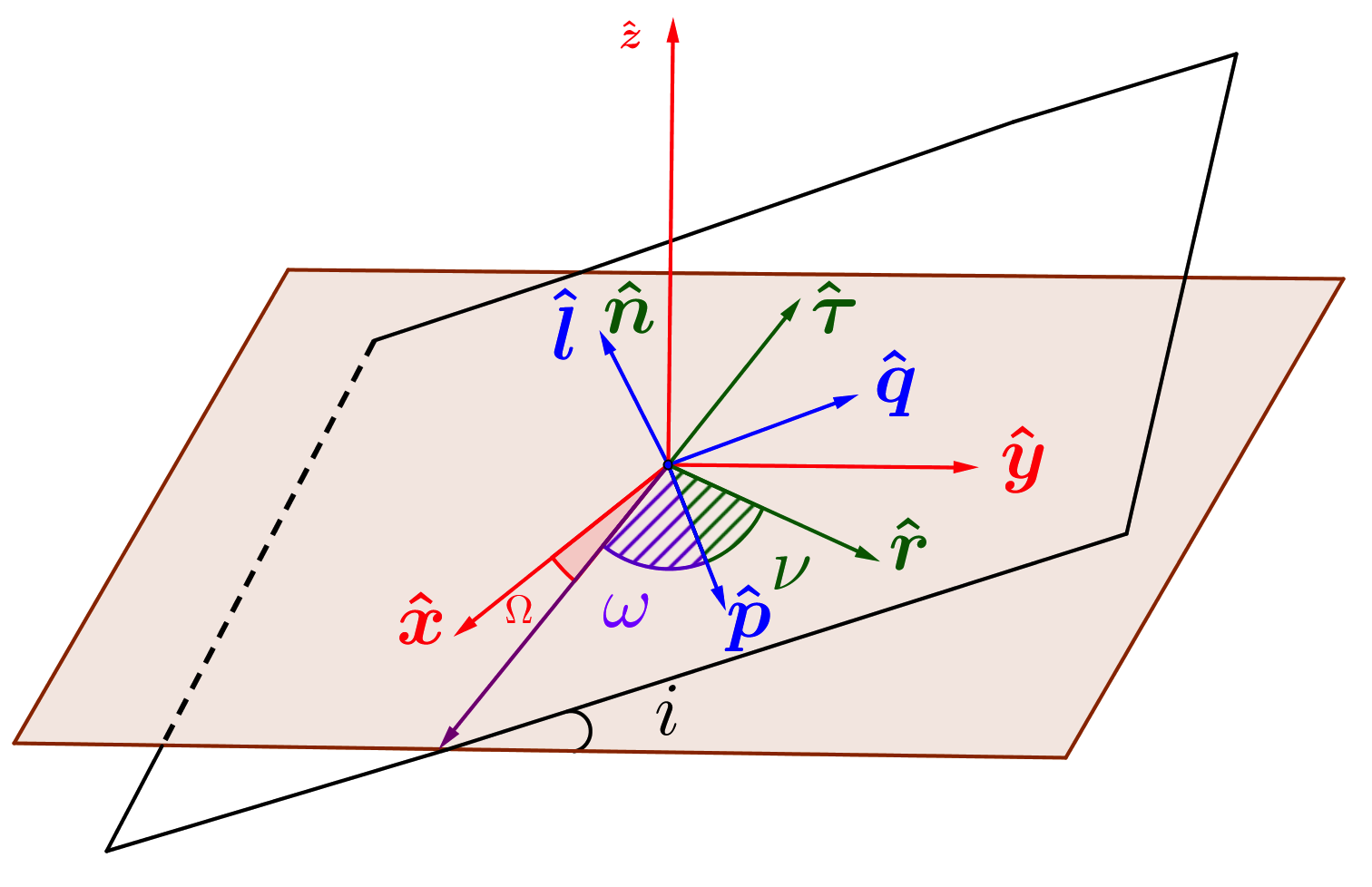}
      \caption{Definition of different reference frames. Solid shaped planes and angles refer to the inertial reference frame $(\bold{\hat{x}},\bold{\hat{y}},\bold{\hat{z}})$, while hatch shaped angles refer to the inclined by angle $i$ orbital plane. In the latter, two reference systems are defined. The reference system $(\bold{\hat{p}},\bold{\hat{q}},\bold{\hat{l}})$ where the unit vector $\bold{\hat{p}}$ points always into the direction of periastron and $\omega$ is the argument of periapsis and the reference system $(\bold{\hat{r}},\bold{\hat{\tau}},\bold{\hat{n}})$ which is rotated by the true anomaly $\nu$ relative to the former.}
\end{figure}

\section{Isotropic wind mass-loss}

The components of a binary system can lose mass in many different ways. Due to the energy and angular momentum conservation laws mass-loss procedures expose the system to additional perturbations including linear and angular momentum recoils. One characteristic way through which stars can lose mass is stellar winds. However, the actual form and structure of these winds is a complicated phenomenon and depends strongly on the structure and rotation of the star as well as on the potential existence of a dynamically important magnetic field in the star. In many cases, e.g., in a supernova explosion, the mass-loss can be assumed to be spherically symmetric. This means that the wind velocity is the same for all points along the surface of the star and thus we refer to this case of wind mass-loss as \emph{isotropic}. We note here that isotropic mass-loss does not produce any momentum kick on the mass-losing body. However, the total mass of the system changes with time and this induces in the system a perturbation which acts as we will see below as a linear drag force. \par
As mentioned above, in the case of \emph{isotropic} wind mass-loss the perturbation $\bold{F}$  in equation $(\ref{aa})$ comes implicitly from the fact that the mass of one or both stars is changing leading to a time-dependent total mass of the system. Figure 4 describes schematically the case of isotropic mass-loss considered in this section. Let's assume that at some point $t$ in time the two-body system has a total mass $M$ and the body $M_{2}$ has exactly the needed velocity $\bold{v}$ to follow the unperturbed orbit (black ellipse in Figure 4). If the total mass of the system momentarily changes by $dM$ introducing no additional momentum kicks to the bodies (what we refer to as isotropic) then the period increases and the velocity $\bold{v}$ is more than needed to follow the unperturbed orbit. The body $M_{2}$ will follow the perturbed orbit instead (red ellipse in FIgure 4). As we will see below we can describe this perturbation as a linear drag force acting to the body $M_{2}$, forcing it to follow the osculating orbit. Before deriving the actual form of the perturbation we remind here that the perturbed orbit is an osculating orbit. This means that the body $M_{2}$ will never follow the osculating orbit since after time $t$ the total mass of the system will continue changing and the actual physical orbit of the body will never be a closed orbit in time. From equations $(\ref{generala})$-$(\ref{generalsigma})$ we know that because of the perturbation induced, the osculating orbit will be both evolved and precessed relative to the unperturbed orbit. In Paper II, we will prove that the osculating orbit as already indicated in Figure 4 has on average the same eccentricity, a greater semi-major axis and it precesses compared to the perturbed orbit. 
\subsection{Perturbing force treatment}
\citet{1963Icar....2..440H} and \citet{1964SvA.....8..127O} independently derived the appropriate perturbation for isotropic wind mass-loss. For a donor of mass $M_{1}(t)$ and a companion of mass $M_{2}(t)$, if the total mass $M(t)=M_{1}(t)+M_{2}(t)$ of the two-body system is changing in time with a rate $\dot{M}(t)<0$, making use of the energy equation leads to the following \emph{perturbation for isotropic wind mass-loss} \citep{1963Icar....2..440H,1964SvA.....8..127O}
\begin{equation}
\bold{F}=-\frac{1}{2}\frac{\dot{M(t)}}{M(t)}\dot{\bold{r}}.\label{isoper}
\end{equation}
\par Perturbation $(\ref{isoper})$ has the form of a linear drag acceleration acting always in the opposite to the motion of the body direction. This linear drag is proportional to the fractional mass-loss rate and is responsible for the reduction of the instantaneous velocity magnitude described above and depicted in Figure 4. This reduction leads to the new osculating orbit which will continue to evolve in time as long as the total mass is continuously changing in time or will be the actual physical orbit of the body if the perturbation ceases instantaneously.
\begin{figure}
  \centering
    \includegraphics[width=0.5\textwidth]{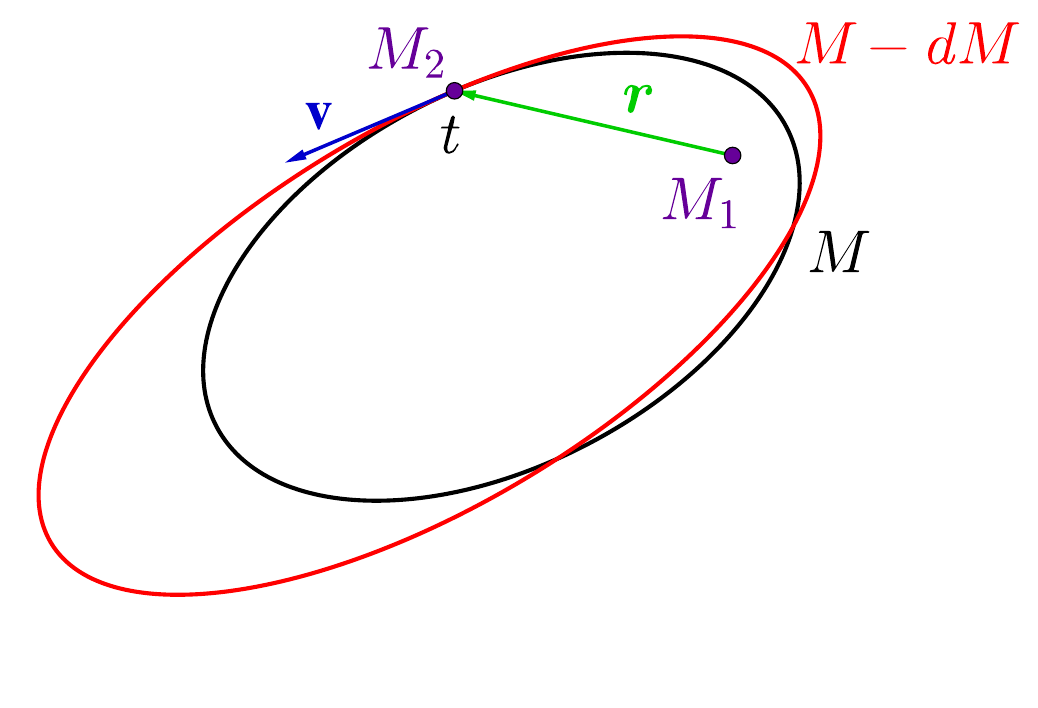}
      \caption{Perturbation induced by instantaneous mass-loss $dM$ at time $t$. The black ellipse is the orbit the body $M_{2}$ would follow if there is no total mass change in the system. The red ellipse is the osculating orbit the body would follow if there is a total mass change $dM$ in the system and the perturbation ceased instantaneously. If there is a mass-loss at time $t$ the velocity $\bold{v}$ is more than needed for the body to follow the black orbit. The linear drag $-\dot{M}/M\bold{\dot{r}}$ induced forces the body $M_{2}$ to follow the osculating red orbit instead.}
\end{figure}
\par Using the perturbing force $(\ref{isoper})$, the equation of motion $(\ref{aa})$ becomes
\begin{equation}\label{eq4}
\ddot{\bold{r}}=-\frac{GM}{r^{3}}\bold{r}-\frac{1}{2}\frac{\dot{M(t)}}{M(t)}\dot{\bold{r}}.
 \end{equation}
\par For the perturbation defined by equation $(\ref{isoper})$, making use of the relations $(\ref{first})-(\ref{last})$ in equations $(\ref{generala})$ and $(\ref{generale})$, one derives the following \emph{phase-dependent time-evolution equations for the osculating semi-major axis and eccentricity in the case of isotropic wind mass-loss} \citep{1963Icar....2..440H} 
\begin{align}
\left(\frac{da}{dt}\right)_{iso}&=-\frac{(1+e^{2}+2e\cos \nu)}{1-e^{2}}\frac{\dot{M}}{M}a\label{isoa}\\
\left(\frac{de}{dt}\right)_{iso}&=-(e+\cos \nu)\frac{\dot{M}}{M}.\label{isoe}
\end{align}
\par Equations $(\ref{isoa})$ and $(\ref{isoe})$ show that the osculating semi-major axis and eccentricity are in general phase-dependent. Using the identity $\cos^{2}\nu + \sin^{2}\nu=1$ we can re-write the nominator in equation $(\ref{isoa})$ as $(e+\cos{\nu})^{2}+\sin^{2}\nu$, which is always positive. This proves that for any isotropic mass-loss the semi-major axis can never decrease throughout the orbit and is not oscillating. Following the same procedure and given that the periastron position is given by $r_{p}=a(1-e)$ one can prove that $\dot{r}_{p}=-\frac{(1-e)(1-\cos\nu)}{1+e}\frac{\dot{M}}{M}a$. Since $e<1$ and $\cos\nu<1$ always for any isotropic mass-loss the periastron position can never decrease throughout the orbit and is not oscillating as well. However, the eccentricity does not follow a similar behaviour and can both increase and decrease throughout the orbit. The eccentricity undergoes oscillations on the orbital timescale \citep[see e.g.,][]{2011MNRAS.417.2104V}. The amplitude of these oscillations is proportional to $\dot{M}/nM$, which is a scaled ratio of the orbital period to the mass-loss timescale. These oscillations can be smoothed out through orbit-averaging procedures when deriving the secular time-evolution equations of the osculating orbital elements. These secular time-evolution equations can be even more simplified in the adiabatic regime [see subsection 2.1.2 and Paper II for more details]. Over time the amplitude of these oscillations is increasing, since eventually the orbital period increases and becomes comparable to the mass-loss time scale. This is the point where adiabaticity is broken and the ``adiabatic" equations no longer apply. 
\subsection{Perturbing Lagrangian treatment}
Equations $(\ref{isoa})$ and $(\ref{isoe})$ can alternatively be derived from equations $(\ref{hugea})$ and $(\ref{hugee})$ for a zero-gauge $(\bold{\Phi}=0)$  by applying the perturbing Lagrangian
\begin{equation}
\Delta L=f(M_{1},M_{2})\left( \frac{1}{2}\dot{\bold{r}}^{2}+\frac{GM}{r}\right)\label{Lagiso}
\end{equation}
where $f(M_{1},M_{2})=\frac{1}{2}\ln{\frac{M}{M_{0}}}$, with $M_{0}$ being a constant equal to the the total mass of the system at $t=0$.
\par We note here that the Lagrangian $(\ref{Lagiso})$ was derived in such a way that it satisfies equation $(\ref{forcelag})$ with the force given by equation $(\ref{isoper})$. Since equation $(\ref{forcelag})$ cannot uniquely determine the relative Lagrangian for a given force, the Lagrangian $(\ref{Lagiso})$ is not unique, has no physical meaning and it acts mainly as a mathematical tool to derive 
the time-evolution of the osculating or non-osculating orbital elements through equations $(\ref{hugea})$  and $(\ref{hugee})$, also allowing for the gauge freedom.  This freedom in the choice of the Lagrangian is depicted in the freedom in the functional dependence of the function $f$ on $M_{1}$ and $M_{2}$. This means that the osculating orbital elements which have a specific physical interpretation cannot depend on the function $f$ and one can prove that equations  $(\ref{hugea})$  and $(\ref{hugee})$ are constructed in such a way that for a zero-gauge that leads to the osculating orbital elements the functional dependence on $f$ diminishes. However, for a non-zero gauge and thus non-osculating orbital elements with no physical meaning this is no longer true. For the generalized Lagrange constraint given in equation $(\ref{nonoscgauge})$ and isotropic wind mass-loss the phase-dependent time-evolution equations for the contact orbital elements [see subsection 2.1.2] are given by
\begin{align}
\left(\frac{da_{con}}{dt}\right)_{iso}&=\frac{f(f+2)}{f+1}\left[\frac{-2nae\sin{\nu}(1+e\cos{\nu})^{2}}{(1-e^{2})^{5/2}}\right]\label{isoacon}\\
\left(\frac{de_{con}}{dt}\right)_{iso}&=\frac{f(f+2)}{f+1}\left[\frac{-nsin{\nu}(1+e\cos{\nu})^{2}}{(1-e^{2})^{3/2}}\right].\label{isoecon}
\end{align}
We note here that equations $(\ref{isoacon})$  and $(\ref{isoecon})$ depend on the arbitrary function $f$ since they describe the time-evolution of the so-called contact elements which as we mentioned before do not have physical meaning. In the case of the non-zero gauge $(\ref{nonoscgauge})$ both the contact semi-major axis and the eccentricity oscillate throughout the orbit. In Paper II we compare the orbit-averaged time-evolution equations for the osculating and contact orbital elements assuming either adiabaticity or delta-function mass-loss at periastron.
\section{Anisotropic wind mass-loss}

Stellar mass-loss is time-dependent and depends in general on the latitude and longitude along the surface of the star. For example, there is much observational evidence that the equatorial regions around rapidly rotating hot-stars generally have an enhanced wind density, perhaps even (e.g., as in Be stars) a circumstellar disk \citep[e.g.,][]{1998Ap&SS.260..149O}. More recent studies assume that the Be-star wind has a disk component in the equatorial plane plus a weak spherical wind above the poles \citep[e.g.,][]{2005ARep...49..709B}. \par The formation of bipolar jets is strongly linked with the accretion onto protostars. Since the accretion process can persist for a long time after the star is born, jets usually accompany a disc \citep[e.g.,][]{1982MNRAS.199..883B} \par
In this section, we do not study the case of a circumstellar disk. However, we assume that the mass-flux rate as well as the wind velocity is different at each point along the spherical surface of the star. We refer here to this case as \emph{anisotropic} wind mass-loss. The results of this section are also applicable to the case of anisotropic planet evaporation (including jets) around the host star in a planet-star binary system \citep[e.g.,][]{2012A&A...537L...3B,2013MNRAS.435.2416V}.\par
\citet{2013MNRAS.435.2416V} were the first to study the case of anisotropic wind mass-loss from the star in a star-planet binary system. However, as we mentioned in Section 3, in this work there are some inconsistencies arising from a mistreated transformation from one reference system to the other. Both the importance of the choice of reference frame and the importance of the use of the right set of equations for the reference system chosen were pointed out in Section 3 as well. For the case of the anisotropic wind mass-loss we describe here, we discuss in Paper II the effect of the corrected time-evolution equations we derived in Section 3 [see equations $(\ref{ref11})$ and  $(\ref{ref12})$] on the secular time-evolution equations of the orbital elements.\par
If the mass-losing body of mass $M_{1}(t)$ has a mass-loss flux per solid angle $J(\phi,\theta,t)>0$, then we can express the mass-loss rate $\dot{M}_{1}$ as
\begin{equation}
\dot{M}_{1}=-\frac{1}{4\pi}\int_{0}^{\pi}\int_{0}^{2\pi}J(\phi,\theta,t)\sin{\theta}d\phi d\theta
\end{equation}
where $\phi$ and $\theta$ are the longitude and co-latitude respectively on the surface of the mass-losing body (donor).\par
If the mass at a specific point is lost with a velocity $\bold{u}(\phi,\theta,t)$ relative to the center-of-mass and along the radius of the donor, then assuming the spin axis of the donor to be fixed with respect to the orbit, the perturbing acceleration $\bold{A}_{aniso}$ arising from the anisotropic wind mass-loss is given by
\begin{equation}
\bold{A}_{aniso}=- \frac{\bold{F}_{aniso}}{M_{1}(t)}\label{acc}
\end{equation}
where $\bold{F}_{aniso}$ is the perturbing force arising from the anisotropic wind mass-loss. In the inertial reference system $K_{I}({\bold{\hat{x}},\bold{\hat{y}},\bold{\hat{z}}})$ mentioned in Section 3 and depicted in Figure 3, the components of this perturbing force are given by 
\begin{scriptsize}
\begin{align}
F_{aniso,x}&=\frac{1}{4\pi}\int_{0}^{\pi}\int_{0}^{2\pi}J(\phi,\theta,t)u(\phi,\theta,t)d\phi \sin{\theta}\cos{\phi}d\theta\label{firstref}\\
 F_{aniso,y}&=\frac{1}{4\pi}\int_{0}^{\pi}\int_{0}^{2\pi}J(\phi,\theta,t)u(\phi,\theta,t)d\phi \sin{\theta}\sin{\phi}d\theta\\
 F_{aniso,z}&=\frac{1}{4\pi}\int_{0}^{\pi}\int_{0}^{2\pi}J(\phi,\theta,t)u(\phi,\theta,t)d\phi \cos{\theta}d\theta .\label{lastref}
\end{align}
\end{scriptsize}
\par It is important to mention here that both the velocity $u(\phi,\theta,t)$ and the mass-flux rate $J(\phi,\theta,t)$ do not depend on the relative position between the two bodies in the binary system. Thus, based on equations $(\ref{acc})-(\ref{lastref})$ one sees that the perturbing force $\bold{F}_{aniso}$ is not phase-dependent. However, both the mass-loss rate and the mass of the donor and thus both the perturbing force and acceleration remain time-dependent. In a similar way to the isotropic wind mass-loss case [see Section 4] where the mass-loss rate and the total mass was a function of time this time dependence induces implicitly a perturbation even though this perturbation is not explicitly phase-dependent. In  Paper II we derive the secular time-evolution equations for the osculating orbital elements in the case of anisotropic wind mass-loss and for the phase-independent perturbing acceleration described by equations $(\ref{acc})-(\ref{lastref})$.
\par Picking a different reference system can lead to equations that are more easy to qualitatively interpret. Given the fact that the wind structure depends on the rotation of the star as well as its magnetic field we usually describe the form of the wind in the inertial reference frame $K_{I}$. However, given the form of the wind in this system we can calculate the inertial perturbing force components and we can then transform for considerable simplifications to the reference system $K_{J}$ using equation $(\ref{trans1})$ or to reference system $K_{R}$ using equation $(\ref{trans2})$. In this new systems the perturbing force components are given respectively by
\begin{equation}
 \begin{pmatrix}
  F_{aniso,p}\\
  F_{aniso,q}\\
  F_{aniso,l} \\
 \end{pmatrix}=
 \begin{pmatrix}
  Q_{11} & Q_{21}  & Q_{31} \\
  Q_{12}  & Q_{22} &  Q_{32} \\
  Q_{13} & Q_{23}  & Q_{33} 
 \end{pmatrix}
  \begin{pmatrix}
  F_{aniso,x}\\
  F_{aniso,y}\\
   F_{aniso,z}\\
 \end{pmatrix}\label{matrixx}
\end{equation}
and
\begin{equation}
 \begin{pmatrix}
  F_{aniso,r}\\
  F_{aniso,\tau}\\
  F_{aniso,n} \\
 \end{pmatrix}=
 \begin{pmatrix}
  \cos{\nu} & \sin{\nu}  & 0\\
  -\sin{\nu}  & \cos{\nu}  &  0 \\
  0& 0  & 1 
 \end{pmatrix}
  \begin{pmatrix}
  F_{aniso,p}\\
  F_{aniso,q}\\
   F_{aniso,l}\\
 \end{pmatrix}.\label{matrix}
\end{equation}
In Paper II we examine qualitatively some simple structures of anisotropic wind mass-loss including jets. We initially begin by describing the wind structure in the inertial reference frame $K_{I}(\bold{\hat{x}},\bold{\hat{y}},\bold{\hat{z}})$ and then by transforming to different reference systems that simplify our equations (reference systems $K_{R}$  and $K_{J}$) we derive the secular time-evolution equations of the orbital elements in these systems in the adiabatic regime.
\section{Non-isotropic ejection/accretion in mass-transfer (reaction-forces)}

Wind mass-loss is only one of the ways two bodies in a binary system can exchange mass. Another possibility is mass ejection or accretion from specific points and with specific velocities. Due to linear momentum conservation, this ejection and accretion induces kicks/reaction-forces to the mass-losing and the mass-accreting star respectively. To distinguish this mass-transfer process from the case of anisotropic wind mass-loss, we choose to refer to this process as \emph{non-isotropic} ejection/accretion. 
\par The most characteristic example of anisotropic ejection/accretion is Roche-Lobe-Overflow (RLOF). In this case,  the mass is ejected from the Lagrangian point $L_{1}$. This mass can $(i)$ escape from the system $(ii)$ be re-accreted from the donor (self-accretion) $(iii)$ hit the radius of the accretor (direct impact) $(iv)$ hit the radius of an accretion disk that has been formed around the companion \citep[e.g.,][]{2007ApJ...667.1170S,2009ApJ...702.1387S,2010ApJ...724..546S}. We refer to the case when all the mass ejected is accreted as conservative mass-transfer while to the case when some mass is lost from the system as non-conservative. 
\par In the next two sections, we study the cases of non-isotropic ejection and accretion in both conservative and non-conservative manner, keeping the ejection/accretion points and velocities as general parameters of the problem.
\subsection{Conservative case ($\dot{M}=\dot{J}_{orb}=0$)}

We assume an eccentric binary system consisting of two masses $M_{1}(t)$ and $M_{2}(t)$. We refer to these bodies as the donor and the accretor, respectively.\par
Binary component $\#1$ loses mass at a rate $\dot{M_{1}}<0$, with absolute velocity $\bold{w}_{1}$ and from point $\bold{r}_{A_{1}}$ relative to the center-of-mass of the body. Binary component $\#2$  accretes mass at a rate $\dot{M_{2}}>0$, absolute velocity $\bold{w}_{2}$ and to point $\bold{r}_{A_{2}}$ relative to the center-of-mass of the body. The mass-transfer stream is assumed to remain always confined to the orbital plane.\par
Following the work of \citet{1963Icar....2..440H} and \citet{2007ApJ...667.1170S}, the equation of motion of the two-body system can be written in an inertial frame as follows
\begin{align}
\nonumber \ddot{\bold{r}}&=-\frac{G(	M	_{1}+M_{2})}{r^{3}}\bold{r}
+\frac{\dot{M_{2}}}{M_{2}}\left( \bold{w}_{2}
+\bold{\omega}_{orb}\times \bold{r}_{A_{2}} \right)\\ 
&-\frac{\dot{M_{1}}}{M_{1}}\left( \bold{w}_{1}
+\bold{\omega}_{orb}\times \bold{r}_{A_{1}} \right)
-\frac{\dot{M_{2}}}{M_{2}} \bold{v}_{2}
+\frac{\dot{M_{1}}}{M_{1}}\bold{v}_{1}\label{dec0}\\ \nonumber
&+\frac{\ddot{M_{2}}}{M_{2}}\bold{r}_{A_{2}}
-\frac{\ddot{M_{1}}}{M_{1}}\bold{r}_{A_{1}}\\
&\nonumber =-\frac{G(	M	_{1}+M_{2})}{r^{3}}\bold{r}+\frac{\dot{M_{2}}}{M_{2}}\left( \bold{w}_{2}-\bold{v}_{2}
+\bold{\omega}_{orb}\times \bold{r}_{A_{2}} \right)\\ \nonumber
&-\frac{\dot{M_{1}}}{M_{1}}\left( \bold{w}_{1}-\bold{v}_{1}
+\bold{\omega}_{orb}\times \bold{r}_{A_{1}} \right)\\
&+\frac{\ddot{M_{2}}}{M_{2}}\bold{r}_{A_{2}}
-\frac{\ddot{M_{1}}}{M_{1}}\bold{r}_{A_{1}}\label{dec1}
\end{align}
where $\bold{v}_{1}$ and $\bold{v}_{2}$ are the orbital velocities of the two bodies with respect to the inertial frame and $\bold{\omega}_{orb}$ is the orbital angular frequency. Equation $(\ref{dec1})$ takes into account the motion of the center-of-mass of the bodies and ignores the forces on the bodies from the mass-transfer stream under the assumption that the stream's mass is negligible. We note here that when writing the linear momentum conservation equations, we assume the absolute velocities $\bold{w}_{1}$ and $\bold{w}_{2}$ to be in the opposite direction to the direction of motion of the bodies.\par
From equation $(\ref{dec1})$ we can see that anisotropic accretion and ejection introduces the reaction forces 
\begin{align}
&\frac{\dot{M_{2}}}{M_{2}}\left( \bold{w}_{2}-\bold{v}_{2}
+\bold{\omega}_{orb}\times \bold{r}_{A_{2}} \right)\label{force1}\\
-&\frac{\dot{M_{1}}}{M_{1}}\left( \bold{w}_{1}-\bold{v}_{1}
+\bold{\omega}_{orb}\times \bold{r}_{A_{1}} \right),\label{force2}
\end{align}
for accretion and ejection respectively. The terms with the second order time derivative in equation $(\ref{dec1})$ are related to the acceleration of the centers-of-mass of the two stars relative to their positions in the unperturbed no mass-loss case [see Section 2 in \citet{2007ApJ...667.1170S} for a detailed description of the relevant terms].\par
The reaction forces $(\ref{force1})$  and $(\ref{force2})$ emerge from applying the linear momentum conservation law in the accretor and the donor respectively. As a reminder, we mention here that in the isotropic mass ejection/accretion case we have 
\begin{align}
&\sum_{i}\left\lbrace \left(\bold{w}_{1}-\bold{v}_{1}\right)_{i}
+\bold{\omega}_{orb}\times \bold{r}_{A_{1i}}\right\rbrace =0\\
&\sum_{i} \left\lbrace\left(\bold{w}_{2}-\bold{v}_{2}\right)_{i}
+\bold{\omega}_{orb}\times \bold{r}_{A_{2i}}\right\rbrace =0.
\end{align}
These terms are zero due to isotropy.
\par If all the mass lost from the donor is accreted by the companion, and any orbital angular momentum transported by the transferred mass is immediately returned to the orbit, then as we mentioned before the mass-transfer is conservative. In this case, both the total system mass and orbital angular momentum are conserved and we can then write $\dot{M_{2}}=-\dot{M_{1}},\ddot{M_{2}}=-\ddot{M_{1}}$. Equations $(\ref{dec0})$ and $(\ref{dec1})$ in the case of conservative mass-transfer yield
\begin{align}
\nonumber \ddot{\bold{r}}&=-\frac{G(	M	_{1}+M_{2})}{r^{3}}\bold{r}\\
&+\dot{M}_{2}\left( \frac{\bold{w}_{2}}{M_{2}}
+\frac{\bold{\omega}_{orb}\times \bold{r}_{A_{2}}}{M_{2}}
+\frac{\bold{w}_{1}}{M_{1}}
+\frac{\bold{\omega}_{orb}\times \bold{r}_{A_{1}}}{M_{1}}
\right)\\ \nonumber
&+\ddot{M}_{2}\left(\frac{\bold{r}_{A_{2}}}{M_{2}}
+\frac{\bold{r}_{A_{1}}}{M_{1}}\right)
-\dot{M}_{2}\left( \frac{\bold{v}_{1}}{M_{1}} +\frac{\bold{v}_{2}}{M_{2}} \right)\\
&=-\frac{G(M	_{1}+M_{2})}{r^{3}}\bold{r}+\bold{b}-\dot{M}_{2}\left( \frac{\bold{v}_{1}}{M_{1}} +\frac{\bold{v}_{2}}{M_{2}} \right)\label{con1}
\end{align}
where we set for convenience 
\begin{align}
\nonumber \bold{b}&\equiv \dot{M}_{2}\left( \frac{\bold{w}_{2}}{M_{2}}
+\frac{\bold{\omega}_{orb}\times \bold{r}_{A_{2}}}{M_{2}}
+\frac{\bold{w}_{1}}{M_{1}}
+\frac{\bold{\omega}_{orb}\times \bold{r}_{A_{1}}}{M_{1}}
\right)\\ 
&+\ddot{M}_{2}\left(\frac{\bold{r}_{A_{2}}}{M_{2}}
+\frac{\bold{r}_{A_{1}}}{M_{1}}\right).\label{beta}
\end{align}
\par Taking into account the center-of-mass of the two bodies motion we can write the positions of the donor and the accretor as
\begin{align}
\label{cm1}\bold{r}_{1}&=\bold{r}_{cm}-\frac{M_{2}}{M_{1}+M_{2}}\bold{r}\\
\label{cm2}\bold{r}_{2}&=\bold{r}_{cm}+\frac{M_{1}}{M_{1}+M_{2}}\bold{r}.
\end{align}
\par The time derivative of the expressions $(\ref{cm1})$ and $(\ref{cm2})$ then gives
\begin{align}
\frac{\bold{v}_{1}}{M_{1}}&=\frac{\bold{v}_{cm}}{M_{1}}-\frac{\dot{M}_{2}}{MM_{1}}\bold{r}-\frac{M_{2}}{M_{1}M}\bold{\dot{r}}\\
\frac{\bold{v}_{2}}{M_{2}}&=\frac{\bold{v}_{cm}}{M_{2}}+\frac{\dot{M}_{1}}{MM_{2}}\bold{r}-\frac{M_{1}}{M_{2}M}\bold{\dot{r}}.
\end{align}
\par In the conservative case we have $\dot{M_{2}}=-\dot{M_{1}},$ and at first order the last term in the right-hand-side (RHS) of equation $(\ref{con1})$ becomes
\begin{equation}
-\dot{M}_{2}\left( \frac{\bold{v}_{1}}{M_{1}} +\frac{\bold{v}_{2}}{M_{2}} \right)=-\dot{M}_{2}\left(\frac{1}{M_{2}}-\frac{1}{M_{1}}\right)\dot{\bold{r}}
\end{equation}
so that the perturbing force in this case can be re-written as
\begin{align}
\bold{F}&=\bold{b}-\dot{M}_{2}\left(\frac{1}{M_{2}}-\frac{1}{M_{1}}\right)\dot{\bold{r}}\\
&=\bold{b}-h\dot{\bold{r}}\label{per1}
\end{align}
where we defined 
\begin{equation}
h\equiv\dot{M}_{2}\left(\frac{1}{M_{2}}-\frac{1}{M_{1}}\right).\label{h}
\end{equation}
\par Applying equations $(\ref{generala})$ and $(\ref{generale})$ with a perturbing force of the form $(\ref{per1})$, using also equations $(\ref{beta})$ and $(\ref{h})$, gives the following \emph{phase-dependent time-evolution equations for the semi-major axis and eccentricity in the conservative mass-transfer case}
\begin{align}
 \nonumber \left(\frac{da}{dt}\right)_{con}&=\frac{2}{n\sqrt{1-e^{2}}}\left[b_{r}e\sin{\nu}
+b_{\tau}(1+e\cos{\nu})\right]\\ 
& -2ha\left[ \frac{e^{2}+2e\cos{\nu}+1}{1-e^{2}}\right]\label{axcon}\\
\begin{split}
\left(\frac{de}{dt}\right)_{con}&=\frac{(1-e^{2})^{1/2}}{na}\left[b_{r}\sin{\nu}
\right. \\
&\left. +b_{\tau}\frac{2\cos{\nu}+e(1+\cos^{2}\nu)}{1+e\cos{\nu}}\right]-2h\left[e+\cos{\nu}\right]\label{econ}
\end{split}
\end{align}
where $b_{r}$ and $b_{\tau}$ are the radial and tangential components of the quantity $\bold{b}$ defined in equation $(\ref{beta})$ and depend on the set up of each problem under consideration. We note here that the quantity $\bold{b}$ does not depend on the relative position between the two bodies in the binary, i.e., it is phase-independent. This is something useful to keep in mind when we perform orbit-averaging processes on equations $(\ref{axcon})$ and $(\ref{econ})$. In Paper II we derive the secular time-evolution equations of the orbital elements for the conservative mass-transfer case in both the adiabatic regime as well as under the assumption of a delta-function mass-loss/transfer at periastron.\par
If one wants to introduce a different gauge, the same equations can be derived from equations  $(\ref{hugea})$  and $(\ref{hugee})$ using the Lagrangian 
\begin{equation}
\Delta L=\bold{b}\bold{r}+f(M_{1},M_{2})\left(\frac{1}{2}\bold{\dot{r}}^{2}+\frac{GM}{r}\right)\label{Lag2}
\end{equation}
where $f(M_{1},M_{2})=\log{\left(\frac{M_{2}}{M_{1}}\right)}-2\log{\left(1+\frac{M_{2}}{M_{1}}\right)}$.
\par It is important to mention that the functional form of $f(M_{1},M_{2})$ does not affect the final evolution equations for the osculating orbital elements, which have physical meaning in the problem. Equations $(\ref{hugea})$ and $(\ref{hugee})$ are constructed in a way that terms that depend on $f(M_{1},M_{2})$ cancel each other out. This is expected since the freedom in the choice of the Lagrangian should not affect the final physical result. In this formulation, the Lagrangian does not have a clear physical interpretation on its own and acts more like a mathematical tool. It remains useful though when one wants to explore the gauge freedom in orbital mechanics.
If one uses the Lagrangian $(\ref{Lag2})$ for \emph{non-osculating} orbital elements that have no physical meaning there is a dependence on $f(M_{1},M_{2})$ in the final equations as we can easily see using the generalized Lagrange constraint defined by equation $(\ref{nonoscgauge})$. Since in this gauge
\begin{equation}
\bold{\Phi}+\frac{\partial\Delta L}{\partial {\bold{\dot{r}}}}=0,
\end{equation}
equations $(\ref{hugea})$ and $(\ref{hugee})$ for the so-called \emph{contact} orbital elements simplify to
\begin{align}
\frac{da_{con}}{dt}&=\frac{2}{na}\left[
\frac{\partial{(-\Delta H^{c})}}{\partial{\sigma}}\right]\label{non1}\\
\nonumber \frac{de_{con}}{dt}&=\frac{1-e^{2}}{na^{2}e}\left[
\frac{\partial{(-\Delta H^{c})}}{\partial{\sigma}}\right]\\
 &-\frac{\sqrt{1-e^{2}}}{na^{2}e}\left[
\frac{\partial{(-\Delta H^{c})}}{\partial{\omega}}\right]\label{non2}
\end{align}
where we defined the non-osculating perturbing Hamiltonian as $\Delta H^{c}$. This Hamiltonian has a different functional dependence on $(\bold{r},\dot{\bold{r}})$ from the relative $\Delta H^{osc}$ computed in the case of osculating orbital elements using a zero gauge $\bold{\Phi}=0$.
\par Using the perturbing Lagrangian $(\ref{Lag2})$ and applying the gauge $(\ref{nonoscgauge})$ into equations $(\ref{non1})$ and $(\ref{non2})$ we have the following phase-dependent time-evolution equations for the so called \emph{contact} orbital elements
\begin{align}
\nonumber \left(\frac{da_{con}}{dt}\right)_{con}&=\frac{2}{n\sqrt{1-e^{2}}}\left[b_{r}e\sin{\nu}
+b_{\tau}(1+e\cos{\nu})\right]\\
&-\frac{2na e\sin{\nu}(1+e\cos{\nu})^{2}}{(1-e^{2})^{5/2}}\left[\frac{f(f+2)}{f+1}\right]\label{non3}\\
\begin{split}
\left(\frac{de_{con}}{dt}\right)_{con}&=\frac{(1-e^{2})^{1/2}}{na}\left[b_{r}\sin{\nu}\right.\\
&\left. +b_{\tau}\frac{2\cos{\nu}+e(1+\cos^{2}\nu)}{1+e\cos{\nu}}\right]\\ 
&-\frac{n\sin{\nu}(1+e\cos{\nu})^{2}}{(1-e^{2})^{3/2}}\left[\frac{f(f+2)}{f+1}\right].\label{non4}
\end{split}
\end{align}
As we noted before equations $(\ref{non3})$ and $(\ref{non4})$ depend on the functional form of the arbitrary function $f$ and thus cannot be attributed any physical meaning. However, as we mentioned in previous sections, in some cases time-evolution equations of non-osculating elements can prove useful and we present them here for completeness. In Paper II we derive and compare the relative secular time-evolution equations for the osculating and contact orbital elements respectively under the assumption of adiabaticity.
\subsection{Non-conservative case, $\dot{M},\dot{J}_{orb}\neq 0$}

In most astrophysical binaries, only a fraction $\gamma$ of the ejected mass is accreted by the companion, i.e., $\dot{M}_{2}=-\gamma \dot{M}_{1}, \dot{M}=(1-\gamma) \dot{M}_{1} $. The rest of the mass is lost from the system carrying along with it some fraction $\zeta >0$ of the total orbital angular momentum. In this case of non-conservative mass-transfer, one should also include the perturbations arising from the loss of total mass and orbital angular momentum.\par
The total angular momentum of the system can change because of three reasons: $(i)$ The velocities of the donor and the accretor change due to mass ejection and accretion respectively $(ii)$ the total system mass changes $(iii)$ the lost mass carries away an excess amount of orbital angular momentum that is not implicitly hidden in mass and velocity change.\par
We have considered the perturbation arising from case $(i)$ in the previous section of conservative mass-transfer. The additional perturbation from cases $(ii)$ and $(iii)$ is described by
\begin{equation}
-\frac{1}{2}(\zeta+1)\frac{\dot{M}}{M}\dot{\bold{r}}.
\end{equation}
where the extra angular momentum carried away by the mass lost from the system is parameterized in terms of the specific angular momentum of the orbit using the parameter $\zeta$ defined as follows
\begin{equation}
\frac{\dot{J}_{orb}}{J_{orb}}=\zeta \frac{\dot{M}}{M}.
\end{equation}
\par The parameter $\zeta$ is a function of time (phase-dependent) in principle depending on both the orbital elements and the stellar properties of the binary components [see second and third paragraph from the end in current section for details].
\par The first term in the parentheses refers to the perturbation from case $(iii)$, expressing the carried-away angular momentum as a fraction of the total orbital angular momentum. The second one refers to the perturbation from case $(ii)$, where we have assumed similar arguments with the ones described for the case of isotropic wind mass-loss in Section 4 and depicted also in Figure 4 [compare also equation $(\ref{isoper})$ in Section 4]. We note here that the form of perturbation will vary if other assumptions about the mass-loss characteristics are made.\par The total perturbation will then be given by
\begin{align}
\bold{F}_{non-con}&=\bold{F}_{con}-\frac{1}{2}(\zeta+1)\frac{\dot{M}}{M}\dot{\bold{r}}\Rightarrow\label{forcenon}\\
\bold{F}_{non-con}&=\bold{b}-h\dot{\bold{r}}-\frac{1}{2}(\zeta+1)\frac{\dot{M}}{M}\dot{\bold{r}}\label{per2}
\end{align}
where the conservative mass-transfer perturbing force is given as described in the previous section by
\begin{align}
\nonumber \bold{F}_{con}&=+\frac{\dot{M_{2}}}{M_{2}}\left( \bold{w}_{2}-\bold{v}_{2}+\bold{\omega}_{orb}\times \bold{r}_{A_{2}} \right)\\
&\nonumber -\frac{\dot{M_{1}}}{M_{1}}\left( \bold{w}_{1}-\bold{v}_{1}
+\bold{\omega}_{orb}\times \bold{r}_{A_{1}} \right)\\
&+\frac{\ddot{M_{2}}}{M_{2}}\bold{r}_{A_{2}}
-\frac{\ddot{M_{1}}}{M_{1}}\bold{r}_{A_{1}}\label{forcenon2}
\end{align}
and $\bold{b}$ and $h$ are defined in the non-conservative case by
\begin{align}
\begin{split}
\bold{b}&\equiv\dot{M}_{2}\left( \frac{\bold{w}_{2}}{M_{2}}
+\frac{\bold{\omega}_{orb}\times \bold{r}_{A_{2}}}{M_{2}}
+\frac{\bold{w}_{1}}{M_{1}}\right.\\ &\left. 
+\frac{\bold{\omega}_{orb}\times \bold{r}_{A_{1}}}{M_{1}}
\right)+\ddot{M}_{2}\left(\frac{\bold{r}_{A_{2}}}{M_{2}}
+\frac{\bold{r}_{A_{1}}}{M_{1}}\right)
\end{split}\\
h&\equiv\dot{M}_{2}\left(\frac{1}{M_{2}}-\frac{1}{M_{1}}\right)\label{h2}\\
\dot{M}_{2}&=\dot{M}_{2,acc}=-\gamma \dot{M}_{1}\label{extrah}\\
\dot{M}_{1}&=-\dot{M}_{2,acc}=\dot{M}_{1,acc}=\gamma \dot{M}_{1}\\
\dot{M}&=(1-\gamma)\dot{M}_{1}.
\end{align}
\par The general perturbation $(\ref{per2})$ now leads to the most general \emph{phase-dependent time-evolution equations for the semi-major axis and eccentricity in the non-conservative mass-transfer case}
\begin{align}
 \nonumber \left(\frac{da}{dt}\right)_{non-con}&=\frac{2}{n\sqrt{1-e^{2}}}\left[b_{r}e\sin{\nu}
+b_{\tau}(1+e\cos{\nu})\right]\\ \nonumber
& -2ha\left[ \frac{e^{2}+2e\cos{\nu}+1}{1-e^{2}}\right]\\
&-(\zeta+1)\left[ \frac{e^{2}+2e\cos{\nu}+1}{1-e^{2}}\right]\frac{\dot{M}}{M}\label{agen}\\
\begin{split}
\left(\frac{de}{dt}\right)_{non-con}&=\frac{(1-e^{2})^{1/2}}{na}\left[b_{r}\sin{\nu}
\right.  \\ 
 &\left. +b_{\tau}\frac{2\cos{\nu}+e(1+\cos^{2}\nu)}{1+e\cos{\nu}}\right]\\
&-2h\left[e+\cos{\nu}\right]-(\zeta+1)\left[e+\cos{\nu}\right]\frac{\dot{M}}{M}.  \label{egen}
\end{split}
\end{align}
\par Equations $(\ref{agen})$ and $(\ref{egen})$ include all the effects on the evolution of the semi-major axis and eccentricity from non-isotropic ejection and accretion in the binary system.\par
The first terms in the RHS of equations $(\ref{agen})$ and $(\ref{egen})$ are related to the absolute velocities of the ejected and accreted matter as well as to the point where the ejection and accretion takes place. This information is hidden in the radial and tangential components of the parameter $\bold{b}$ defined by equation $(\ref{beta})$. We mention here that the knowledge of the final accretion points and velocities given the initial ones requires the full solution of the two-body plus a mass-transfer stream problem. Applications have been undertaken to approximate the mass-transfer stream as a third point particle and numerically integrate the equations of motion assuming ballistic trajectories \citep[e.g.,][]{2010ApJ...724..546S,2014A&A...570A..25D}. \par
The second terms in the RHS of equations $(\ref{agen})$ and $(\ref{egen})$ refer to the mass that is accreted. The accreted mass is parametrized with $\gamma$ defined in equation $(\ref{extrah})$ while the perturbation has the form of $h$ defined in equation $(\ref{h2})$. The fraction of accreted mass $\gamma$ is a free parameter in a problem but can be constrained using ballistic or hydrodynamic simulations. Furthermore, for phase-dependent mass-transfer the accreted mass can also be confined following e.g., the Bondi-Hoyle accretion scenario. The case of phase-dependent mass-transfer will be discussed in detail in Paper II.\par
The last terms in equations $(\ref{agen})$ and $(\ref{egen})$ refer to the mass lost from the system parametrized by $\gamma$ as well as the orbital angular momentum this mass carries with it parametrized by $\zeta$. The second term inside the square brackets refers to the former while the first one refers to the latter. 
\par We note here that the $\zeta$-dependent term refers to the additional angular momentum removed by the lost mass, i.e., to the change in the total orbital angular momentum rather than the one due to the total system mass change. The sources of this additional angular momentum are many. They include the extra angular momentum the mass-transfer stream gains when moving from the donor to the accretor and exchanges of angular momentum between the orbit and  the stellar spins due to tidal torques. We express here this amount of angular momentum removal as a part of the total orbital angular momentum. This is strongly dependent on the specific assumptions made in the set up of each problem. For example, if we assume that the total mass lost from the system occurs in the vicinity of the accretor, carrying along with it the specific angular momentum of the latter then we have $\zeta=M_{1}/M_{2}$ \citep[e.g.,][]{1997A&A...327..620S,2009ApJ...702.1387S}. In general, $\zeta$ is a free parameter and its value can be restricted using three-body ballistic trajectories \citep[e.g.,][]{1975MNRAS.170..325F,1993ApJ...410..719B} and Appendix A.3 of \citet{2013ApJ...764..166D} for a review. Another approach to limit some of the aforementioned free parameters incuding mass-transfer rates is to study mass-transfer with SPH simulations \citep[e.g.,][]{2005MNRAS.358..544R,2009MNRAS.395.1127C,2011ApJ...726...67L}. 
\par We note here that in the limiting case of $\zeta \rightarrow 0$ (no extra angular momentum removed from the system) the last terms in the RHS of equations $(\ref{agen})$ and $(\ref{egen})$ become equivalent to the terms in the RHS of equations $(\ref{isoa})$ and  $(\ref{isoe})$ presented in Section 4 in the case of isotropic wind mass-loss. This emerges from the fact that when there is no extra angular momentum carried away from the system the orbital evolution induced by the change in the total orbital momentum is implicitly induced by the total mass change with time. Any extra angular momentum removed from the system is parametrized by $\zeta \neq 0$ and induces an additional orbital evolution of the system while this effect is maximized in the limiting case of $\zeta \rightarrow 1$.
\par In this Section we presented the form of the phase-dependent time-evolution equations of the orbital elements in the most general case of non-conservative mass-transfer assuming extended bodies. In Paper II we orbit-average these equations and derive the secular time-evolution equations of the orbital elements for the non-conservative case in the adiabatic regime as well as under delta-function mass-loss/transfer at periastron.
\section{Point-mass approximation}

The validity of any model describing mass-loss and mass-transfer in a binary system (e.g., the Roche model) relies on two assumptions: that the stellar components can be treated as point masses and that the donor star is rotating synchronously with the orbit. For the latter approximation, we assume that the bodies are synchronized with the orbit. For the former approximation we present here the form the perturbing force and the time-evolution equations of the orbital-elements reduce to in the case of point-masses.\par
Under the point-mass approximation, the positions of the ejection and accretion points relative to the center-of-mass of the bodies is zero and so we have $\bold{r}_{A1}=\bold{r}_{A2}=0$. Using equations $(\ref{forcenon})$ and $(\ref{forcenon2})$, the perturbing force in this case is reduced to
\begin{align}
\nonumber \bold{F}_{non-con}&=\frac{\dot{M_{2}}}{M_{2}}\left( \bold{w}_{2}-\bold{v}_{2}\right)
-\frac{\dot{M_{1}}}{M_{1}}\left( \bold{w}_{1}-\bold{v}_{1}
 \right)\\
 &-\frac{1}{2}(\zeta+1)\frac{\dot{M}}{M}\dot{\bold{r}}\\
 &=\bold{c}-k\dot{\bold{r}}\label{point0}
\end{align}
where we defined 
\begin{equation}
\bold{c}\equiv\frac{\dot{M_{2}}}{M_{2}}\left( \bold{w}_{2}-\bold{v}_{2}\right)
-\frac{\dot{M_{1}}}{M_{1}}\left( \bold{w}_{1}-\bold{v}_{1}
 \right),\label{point1}
\end{equation}
 and
 \begin{equation}
 k\equiv\frac{1}{2}(\zeta+1)\frac{\dot{M}}{M}.\label{point2}
\end{equation}
\par Making use of equations $(\ref{point0})-(\ref{point2})$, equations $(\ref{agen})$ and $(\ref{egen})$ give the \emph{phase-dependent time-evolution equations of the semi-major axis and eccentricity for non-conservative mass-transfer and in the case of point masses} 
\begin{align}
 \nonumber \left(\frac{da}{dt}\right)_{non-con}^{point}&=\frac{2}{n\sqrt{1-e^{2}}}\left[c_{r}e\sin{\nu}
+c_{\tau}(1+e\cos{\nu})\right]\\
&-(\zeta+1)\left[ \frac{e^{2}+2e\cos{\nu}+1}{1-e^{2}}\right]\frac{\dot{M}}{M}\label{agenp}\\
\begin{split}
\left(\frac{de}{dt}\right)_{non-con}^{point}&=\frac{(1-e^{2})^{1/2}}{na}\left[c_{r}\sin{\nu}
\right. \\
&\left. +c_{\tau}\frac{2\cos{\nu}+e(1+\cos^{2}\nu)}{1+e\cos{\nu}}\right]\\
&-(\zeta+1)\left[e+\cos{\nu}\right]\frac{\dot{M}}{M}.\label{egenp}
\end{split}
\end{align}
where now $c_{r}$ and $c_{\tau}$ are the radial and tangential components respectively of the quantity $\bold{c}$ defined in equation $(\ref{point1})$. These components depend only on the relative velocity of the ejected or accreted matter to the motion of the mass-loosing or accreting body.
\par In this Section we derived the phase-dependent time-evolution equations of the orbital elements for non-conservative mass-transfer assuming point masses. In Paper II we orbit-average these equations and derive the secular time-evolution equations of the orbital elements under the point-mass approximation and for some commonly used characteristic examples of mass-loss/transfer in the adiabatic regime.
\section{Conclusions}

Motivated by observations of many mass-transferring eccentric binary systems, in this paper we derive self-consistent equations for the orbital evolution of interacting eccentric binaries.\par
We presented a general formalism to derive the time-evolution equations of the orbital elements of a binary exposed to general perturbations. We applied this formalism to the case of mass-loss and mass-transfer in a binary, treating the latter as perturbations to the general two-body problem. We first discussed the cases of isotropic and anisotropic wind mass-loss. We then studied the general case of non-isotropic ejection and accretion in conservative as well as non-conservative manner for both point masses and extended  bodies.
\par The main results of the paper are summarized below:
\begin{itemize}
\item We discussed the physical interpretation of the osculating orbital elements within the context of gauge freedom in celestial mechanics. We presented two different ways to derive the phase-dependent time-evolution equations of the orbital elements and commented on the advantages of each method.
\item We pointed out the importance of the choice of reference frame since an incorrect treatment can lead to misinterpretations of the orbital evolution of the system. We derived the phase-dependent time-evolution equations of the semi-major axis and eccentricity in three different reference frames pointing out the importance of using the right set of time-evolution equations for the chosen reference frame. Comparing with other work in the literature we presented the necessary corrections that have to be made so that the chosen set of time-evolution equations of the orbital elements is consistent with the chosen reference frame in the problem.
\item We studied for completeness the case of isotropic wind mass-loss. We verified that for isotropic mass-loss, the semi-major axis as well as the periastron position can never decrease throughout the orbit. On the contrary, the eccentricity undergoes oscillations on orbital time-scales. The amplitude of these oscillations is proportional to the ratio of the orbital period to the mass-loss timescale and increases over time.
\item We considered the case of anisotropic wind mass-loss. Under the assumption of phase-independent anisotropic wind mass-loss the orbital evolution depends on the velocity and structure of the stellar wind along the surface and relative to the center-of-mass of the mass-losing body.  By making use of the correct transformation treatment between different reference frames that we described in a previous section, we have presented the correct set of time-evolution equations of the orbital elements that are necessary to study the case of anisotropic wind mass-loss.
\item For either the case of conservative or non-conservative mass-transfer for extended bodies we derived the phase-dependent time-evolution equations of the osculating orbital elements. The mass-transfer effect on the semi-major axis and eccentricity depends strongly on the binary characteristics and is proportional to the mass-loss and mass accretion rates. The mass-transfer may either increase or decrease the eccentricity in contrast to tides which always act to circularize the orbit.  In the case of extended bodies, the relative ejection and accretion points and associated velocities play an important role as well. Furthermore, in the case of non-conservative mass-transfer there is a strong dependence on the fraction of the total orbital angular momentum the lost mass removes from the system, which in all cases remains a free parameter.
\item The point-mass assumption is often the first approximation to be made in problems regarding mass-transfer in binary systems. We presented the form of the time-evolution equations of the orbital elements in the non-conservative non-isotropic ejection and accretion case for point masses. We concluded by pointing out the remaining dependence on the relative velocity of the ejected/accreted mass to the motion of the mass-losing/accreting body.
\end{itemize}

In paper II, we orbit-average the general phase-dependent time-evolution equations of the orbital elements presented in this paper. We do that under the assumption of either adiabaticity or delta-function mass-loss/transfer at periastron and for all the different types of mass-loss/transfer studied here. The secular (orbit-averaged) time-evolution equations for the semi-major axis and eccentricity are useful analytical equations that can be used to model the evolution of interacting eccentric binaries in star and binary evolution codes like StarTrack \citep{2008ApJS..174..223B}, BSE \citep{2002MNRAS.329..897H} or MESA \citep{2011ApJS..192....3P,2013ApJS..208....4P,2015ApJS..220...15P}.

\bigskip
\section*{ACKNOWLEDGEMENTS}
We want to thank our colleagues Dimitri Vears, Jeremy Sepinsky and Lorenzo Iorio who provided insight and expertise that greatly assisted the research. We are thankful to their comments that greatly improved the manuscript during its composition and revising stages.

 
\newpage
\appendix
\section{Appendix A}

Using the six Keplerian elements $(a,e,i,\Omega,\omega,\sigma)$ defined in Section 2.1.1,  we calculate the partial derivatives of the unperturbed position $\bold{f}$ and velocity $\bold{g}$ with respect to the orbital elements and choosing the reference system $K_{R}(\hat{\bold{r}},\hat{\bold{\tau}},\hat{\bold{n}})$ defined in Section 4 and depicted in Figure 3. The unperturbed position and velocity as well as their various partial derivatives with respect to the orbital elements are given as functions of the orbital elements and the true anomaly $\nu$ and can be written as follows
\begin{align}
\left(\bold{f}\right)_{r}&=\frac{a(1-e^{2})}{1+e\cos \nu}\label{first}\\
\left(\bold{f}\right)_{\tau}&=0\\
\left(\bold{f}\right)_{n}&=0\\
\left(\bold{g}\right)_{r}&=\frac{na}{\sqrt{1-e^{2}}}e\:\sin \nu\\
\left(\bold{g}\right)_{\tau}&=\frac{na}{\sqrt{1-e^{2}}}(1+e\: \cos \nu)\\
\left(\bold{g}\right)_{n}&=0\\
\left(\frac{\partial{\bold{f}}}{\partial{\sigma}}\right)_{r}&=
\frac{a}{\sqrt{1-e^{2}}}e\: \sin \nu\\
\left(\frac{\partial{\bold{f}}}{\partial{\sigma}}\right)_{\tau}&=
\frac{a}{\sqrt{1-e^{2}}}(1+e\: \cos \nu)\\
\left(\frac{\partial{\bold{f}}}{\partial{\sigma}}\right)_{n}&=0\\
\left(\frac{\partial{\bold{f}}}{\partial{\omega}}\right)_{r}&=0\\
\left(\frac{\partial{\bold{f}}}{\partial{\omega}}\right)_{\tau}&=
\frac{a(1-e^{2})}{1+e\cos \nu}\\
\left(\frac{\partial{\bold{f}}}{\partial{\omega}}\right)_{n}&=0\\
\left(\frac{\partial{\bold{g}}}{\partial{\sigma}}\right)_{r}&=
-\frac{na}{(1-e^{2})^{5/2}}(1+e\:\cos \nu)^{2}\\
\left(\frac{\partial{\bold{g}}}{\partial{\sigma}}\right)_{\tau}&=0\\
\left(\frac{\partial{\bold{g}}}{\partial{\sigma}}\right)_{n}&=0\\
\left(\frac{\partial{\bold{g}}}{\partial{\omega}}\right)_{r}&=
-\frac{na}{\sqrt{1-e^{2}}}(1+e\:\cos \nu)\\
\left(\frac{\partial{\bold{g}}}{\partial{\omega}}\right)_{\tau}&=
\frac{na}{\sqrt{1-e^{2}}}(e\:\sin \nu)\\
\left(\frac{\partial{\bold{g}}}{\partial{\omega}}\right)_{n}&=0\\
\left(\frac{\partial{\bold{g}}}{\partial{a}}\right)_{r}&=
-\frac{n}{\sqrt{1-e^{2}}}\sin \nu\\
\left(\frac{\partial{\bold{g}}}{\partial{a}}\right)_{\tau}&=
\frac{n}{\sqrt{1-e^{2}}}(1+\:\cos \nu)\\
\left(\frac{\partial{\bold{g}}}{\partial{a}}\right)_{n}&=0.\label{last}
\end{align}

\end{document}